\documentclass[11pt]{article}
\usepackage[cp1251]{inputenc}
\usepackage{cyr}
\usepackage{epsf}
\usepackage{pazh}
\usepackage[russian]{babel}
\usepackage{morefloats}
\tightenlines
\def\*{$^{*}$}
\def\Б{$^{\mbox{\small Б}}$}
\def\В{$^{\mbox{\small В}}$}
\def\Ч{$^{\mbox{\small Ч}}$}
\def\З{$^{\mbox{\small З}}$}
\def\Д{$^{\mbox{\small Д}}$}
\def\ЕТЗУ{ЬТЗ~У$^{-1}$}
\def\ЕТЗУН{ЬТЗ~УН$^{-2}$~У$^{-1}$}
\def\etal{{et~al.}}
\def\t90{{Т$_{90}$}}
\def\cm2{{cm$^{2}$}}
\def\m2{{m$^{2}$}}
\def\gr{{$^{\circ}$}}
\def\si{{$\sim$}}

\sloppypar

\begin{document}
\baselineskip 21pt

\title{\bf Catalog of Short Gamma-Ray Transients Detected in the SPI/INTEGRAL Experiment}

\author{\bf \hspace{-1.3cm} \ \
P.~Yu. Minaev\affilmark{1*}, A.~S. Pozanenko\affilmark{1}, S.~V. Molkov\affilmark{1}, S.~A. Grebenev\affilmark{1}}

\affil{
{\it $^1$ Space Research Institute, Russian Academy of Sciences}}

\vspace{2mm}
%\received{~~~~~~~~}
%\received{\today}
%\revised{}

\sloppypar \vspace{2mm} \noindent

We analyzed data obtained by the SPI telescope onboard the INTEGRAL observatory to search for short transient events with a duration from 1 ms to a few tens of seconds. An algorithm for identifying gamma-ray events against the background of a large number of charged particle interactions with the detector has been developed. The classification of events was made. Apart from the events associated with cosmic gamma-ray bursts (GRBs) confirmed by other space experiments and the activity of known soft gamma repeaters (for example, SGR 1806-20), previously unreported GRBs have been found. GRB candidates and short gamma-ray events probably associated with the activity of known SGRs and AXPs have been selected. The spectral evolution of 28 bright GRBs from the catalog has been studied extensively. A new method for investigating the spectral evolution is proposed. The energy dependence of the spectral lag for bursts with a simple structure of their light curves and for
individual pulses of multipulse events is shown to be described by a logarithmic function, $lag \sim A\log(E)$. It has been established that the parameter A depends on the pulse duration, with the dependence being universal for all of the investigated GRBs. No negative spectral lags have been detected for bursts with a simple structure of their light curves.

\noindent {\bf keywords:\/} gamma-ray transients, gamma-ray bursts, soft gamma repeaters, anomalous X-ray pulsars, spectral evolution, spectral lag, cross-correlation analysis

\noindent
%{\bf PACS codes:\/} 95.55.-n, 95.75.-z, 95.80.+p, 95.85.Nv, 95.85.Pw, 98.70.-f

\vfill
\noindent\rule{8cm}{1pt}\\
{$^*$ $<$minaevp@mail.ru$>$}

\clearpage

%***************************************************************
\section*{\textbf{INTRODUCTION}}
\noindent

Gamma-ray bursts (GRBs) are among the most powerful explosions in the Universe. Despite the fact that they have been actively studied for almost half a century, their nature is not yet clear.

GRBs are commonly classified into two types associated with two different progenitors. The first type, short GRBs, is probably associated with the merging of compact components (neutron stars, black holes) in a binary system; as a consequence, this class of GRBs is also called mergers (Paczynski 1986; Meszaros and Rees 1997, 1992; Rosswog and Ramirez-Ruiz 2003; Rosswog et al. 2003). Most short GRBs have a duration \t90 less than 2 s (Kouveliotou et al. 1993; Norris et al. 2005). The characteristic duration that separates the long bursts from the short ones depends on the spectral range (see, e.g., Minaev et al. 2010b). The phenomenological properties of short GRBs are considered in detail in Donaghy et al. (2006). However, an extended emission with a duration of tens of seconds was detected both in the light curves of individual events and in the total light curve of a group of events (Lazzati et al. 2001; Connaughton 2002; Frederiks et al. 2004; Montanari et al. 2005; Gehrels et al. 2006; Minaev et al. 2009, 2010a, 2010b). The nature and models of the extended emission are discussed, for example, in Metzger et al. (2008) and Barkov and Pozanenko (2011). It has not yet been ascertained whether the extended emission is a common property of all short GRBs.

The second type of GRBs, long bursts, is probably associated with the core collapse of supermassive stars; therefore, this class is also called collapsars (Woosley 1993; Paczynski 1998; Fryer et al. 1999; Meszaros 2006). In some cases, a long GRBs is also accompanied by the observation of a type Ib/c supernova (Galama et al. 1998; Hjorth et al. 2003), which is often called a hypernova (Paczynski 1998) due to its very high luminosity that exceeds the luminosity of typical supernovae of this type by several orders of magnitude (Kulkarni et al. 1998). The duration of most events of this class is longer than 2 s (Kouveliotou et al. 1993; Norris et al. 2005).

The fraction of short bursts is 25\% in the BATSE experiment (Kouveliotou et al. 1993) and only 4\% in the IBIS/ISGRI experiment onboard the INTEGRAL observatory (Vianello et al. 2009). The dependence of the fraction of short GRBs on the lower energy threshold of the GRB detector trigger is discussed in Minaev et al. (2010b). This dependence shows that the fraction of short bursts increases with increasing lower energy threshold of the detector trigger. This dependence stems from the fact that short GRBs have a harder spectrum than long ones and, therefore, they are easier to detect in a harder energy band. Indeed, the lower IBIS/ISGRI trigger threshold is lower than the BATSE threshold and, therefore, the fraction of short GRBs in the IBIS/ISGRI experiment can be lower than that in the BATSE experiment. The expected value is about 10\% (the observed one is 4\%). Thus, a deficit of short GRBs is observed in the IBIS/ISGRI experiment even if the dependence of the fraction of short GRBs on the lower energy threshold of the detector sensitivity is taken into account.

As was shown by Minaev et al. (2009, 2010a, 2010b), the fraction of the short GRBs registered by the anticoincidence shield (ACS) of the SPI spectrometer onboard the INTEGRAL observatory can reach 45\%, which is considerably larger than has been thought up until now. However, the fraction of unconfirmed short events in SPI–ACS is 30\%. Thus, on the one hand, a deficit of short GRBs is observed in the IBIS/ISGRI experiment and, on the other hand, an excess of short unconfirmed events is observed in the SPI-ACS experiment (see also Rau et al. 2005).

The INTEGRAL Burst Alert System (IBAS) is used only to analyze the IBIS/ISGRI and SPI–ACS data (Mereghetti et al. 2003) and is not used to analyze the SPI data. The SPI spectrometer is more sensitive to hard gamma-ray emission with an energy above 100 keV (compared to IBIS/ISGRI) which is typical of short GRBs. Therefore, gamma-ray events with a hard energy spectrum previously undetected by IBAS, including short GRBs, can be present in the SPI/INTEGRAL data. One of our goals is the search for short GRBs in the SPI/INTEGRAL data in a wide energy range, (20, 8000) keV. The search for transient events in the SPI data in the energy range corresponding to the 511-keV annihilation line was carried out by Tsygankov and Churazov (2010).

Apart from the duration and spectral hardness, the spectral lag, a parameter characterizing the spectral evolution, is used in GRB studies. The lag is the time shift of the light curve profiles in different energy channels and is considered positive if the time profile in softer energy channels “lags behind” that in hard ones. A small positive lag is characteristic of short GRBs, while there is no lag within the statistical error limits in some of the bursts (Norris and Bonnell 2006;
Zhang et al. 2006). Larger positive spectral lags are characteristic of long GRBs. An empirical dependence of the lag on luminosity has been detected for the class of long GRBs (Norris et al. 2000; Norris 2002; Gehrels et al. 2006; Schaefer 2007; Hakkila et al. 2008; Ukwatta et al. 2012).

There are several theoretical models that explain the nature of the spectral lag. One of the models discussed in Dermer (1998), Kocevski and Liang (2003), and Ryde (2005) suggests the spectral evolution during the burst central engine activity (prompt emission) related to a change in the properties of the burst source itself. Peng et al. (2011) hypothesized that the spectral evolution of the burst central engine could explain both positive and negative lags under certain conditions.

Another model that explains the nature of the lags is based on the relativistic kinematic effect (curvature effect) and is discussed in Salmonson (2000), Ioka and Nakamura (2001), Dermer (2004), Shen et al. (2005), Lu et al. (2006), and Ukwatta et al. (2012). If the photons are assumed to be emitted from a spherically symmetric, relativistically expanding shell simultaneously in the comoving frame, then an observer at infinite distance will record the photons emitted along the source–observer axis earlier than the off-axis photons. In this case, the observed photon energy depends on the angle between the photon propagation direction and the source–observer axis: the larger the angle, the “softer” the photon, i.e., the photons emitted along the source–observer axis are recorded earlier and have the highest energy. Thus, the kinematic model explains only the positive lags. A situation where both the kinematic effects and the change in central engine properties are responsible for the lags is also possible (Peng et al. 2011). The kinematic effects always increase the lag, while the second mechanism can explain both negative and positive lags.

There is one more possible model that describes the negative lags. As will be considered below, the negative lags can result from the superposition of individual pulses, with the lag for each pulse being positive and the spectral hardness of the pulses increasing with time (from pulse to pulse).

Hakkila and Preece (2011) showed that the observed properties of short pulses in long GRBs with a complex multipulse structure are similar to those of short GRBs. This may imply that, despite the differences in many observed properties of short and long bursts, there is a common physical mechanism for both types.

Here, we search for and investigate GRBs and other short gamma-ray transients in the SPI/INTEGRAL data. We have compiled a catalog of short gamma-ray transients registered by the SPI/INTEGRAL experiment. We also investigate the spectral evolution of GRBs based on the SPI and IBIS/ISGRI data. We have described a new method for investigating the spectral evolution, suggested a parameter characterizing the spectral evolution, and detected peculiarities of the spectral evolution of events with a complex structure of their light curves.

\section*{\textbf{THE CATALOG OF SHORT GAMMA-RAY TRANSIENTS OF THE SPI/INTEGRAL EXPERIMENT}}
\subsection*{\textbf{The SPI, SPI-ACS, and IBIS/ISGRI Experiments}} \noindent

The INTEGRAL observatory was launched into a highly elliptical orbit (the perigee and apogee of its initial orbit are 9000 and 153 000 km, respectively) with a period of 72 hours on October 17, 2002 (Jensen et al. 2003). The observatory consists of the IBIS/ISGRI, SPI, JEM-X, OMC telescopes and the SPI anticoincidence shield (SPI–ACS). All of the aperture telescopes (SPI, IBIS/ISGRI, JEM-X, OMC) onboard the observatory are aligned.

IBIS is a coded-mask gamma-ray telescope (Ubertini et al. 2003). Its energy range is from 15 keV to 10 MeV. The photons in the energy ranges (15, 300) keV and (0.3, 10) MeV are registered by the ISGRI (CdTl elements) and PICsIT (CsI elements) detectors, respectively. The telescope's detectors have a total geometrical area of \si 2500 \cm2. The field of view is 30\gr at the zero sensitivity level.

The SPI gamma-ray spectrometer is an array of 19 hexagonal detectors made of ultra-pure germanium with a total geometrical area of 508 \cm2 (Vedrenne et al. 2003). The spectral resolution of the SPI/INTEGRAL spectrometer is 2.2 keV at 1.33 MeV. The energy range is 20 keV - 8 MeV. A coded mask made of tungsten is used for imaging. The telescope's total field of view is 30\gr. It should be noted that the SPI field of view is slightly wider than the IBIS/ISGRI one and has a different shape (hexagonal versus square). Therefore, most of the events observed in the SPI experiment are also present in the IBIS/ISGRI data, with the exception of several events located at the edge of the SPI field of view.

SPI-ACS composed of 91 bismuth germanate (BGO) crystals with a maximum effective area of 0.3 \m2 (von Kienlin et al. 2003) is used to increase the sensitivity of the SPI telescope by removing the background associated with the interaction of the observatory with cosmic rays. Each BGO crystal is viewed by two photomultipliers (PMTs) and the counts from all PMTs are recorded in a single channel. SPI-ACS records photons from almost all directions. The direction coincident with the SPI field of view is least sensitive. SPI-ACS has a lower sensitivity threshold of \si 80 keV - the physical properties of the individual BGO assemblies (detector + PMT + discriminator) slightly differ and, therefore, they have different lower energy thresholds: from 60 to 120 keV; the upper energy threshold is \si 10 MeV. The SPI-ACS
time resolution is 50 ms (von Kienlin et al. 2003). IBAS (INTEGRAL Burst Alert System) (Mereghetti et al. 2003) is used to analyze the SPI-ACS data. The IBAS software algorithm identifies events on nine different time scales (0.05, 0.1, 0.2, 0.4, 0.8, 1, 2, and 5 s) provided that the event significance with respect to the mean background is 9, 6, 9, 6, 9, 6, 9, and 6$\sigma$, respectively. The light curves of the events identified by this algorithm (containing data from -5 to 100 s relative to the trigger time) are publicly accessible (http://isdcarc.unige.ch/arc/FTP/ibas/spiacs/). IBAS is also used in an automatic analysis of the IBIS/ISGRI data, but IBAS is not used simultaneously with the SPI spectrometer.

\subsection*{\textbf{The Data Processing Algorithm}}

We analyzed the data obtained in the SPI, SPI–ACS, and IBIS/ISGRI experiments over the period from July 12, 2003 to January 23, 2010 (91–888 INTEGRAL revolutions). The revolutions from 1 to 90 were excluded from our processing, because of various calibrations were made at the beginning of the INTEGRAL mission, also the data recording and processing codes were modified, and the background in the detectors was unstable.

\textbf{Our main goal was the search for short GRBs in the SPI data}. The SPI–ACS and IBIS/ISGRI data were used only for a joint analysis of the events detected initially in the SPI data.

We used the SPI data processing algorithm that select trigger events on time scales of 0.001, 0.01, 0.1, 1, and 10 s with thresholds of 20, 6, 5, 5, and 4$\sigma$, respectively. The 20$\sigma$ threshold for the 0.001 s interval was chosen to minimize the number of fluctuations over the investigated period of
time so as to obtain only the high-intensity events during the initial selection. Only very intensive short events with a duration of 0.001 s could be confirmed by other experiments (with a lower time resolution). The charged particle interactions with the detector constitute an overwhelming majority of such trigger events. The thresholds for the remaining time scales were chosen so as not to lose the low-intensity events. Despite the fact that the estimated number of fluctuations over the investigated period is large (500 or more), the total number of triggered events admits a further selection according to criteria 1–6 (see below).

The sum of three types of counts was used in analyzing the SPI data: SGL is the ''ordinary'' count recorded in a single detector; PSD is the count recorded in a single detector whose pulse shape confirms its photon nature; and DBL are the counts recorded simultaneously in two different detectors due to the Compton scattering of an initial photon inside one of the detectors.

The significance of the \textit{i}th event, $\sigma$$_{i}$, was determined in the energy range (20, 650) keV and calculated from the formula
\begin{eqnarray}
\sigma_{i} = \frac{C_{i}-F_{k}}{\sqrt{F_{k}}}
\label{one}.
\end{eqnarray}
where C$_{i}$ is the number of counts in the investigated \textit{i}th bin of the light curve (the sum of the counts from all detectors), F$_{k}$ is the background level defined as the mean number of counts in the \textit{k}th interval (-50, 0) s relative to the start time of the investigated \textit{i}th bin. The counts with an energy above 650 keV were excluded from this step of our analysis, because some of the counts in this range are associated with the electronic noise. In the investigated time interval (2003–2010), the number of operational SPI detectors decreased from 19 to 17.

A total of more than one hundred thousand events were selected by the algorithm on various time scales (from 1 ms to 10 s). For each detected event, we constructed its light curve and energy–time diagram and analyzed the distribution of photons over the detectors plane.

Three classes of events were formed: fluctuations, candidates for ''real'' gamma-ray events (for example, GRBs), and three types of instrumental events associated with the interaction of the detector with charged particles (the interactions with electron beams, protons, and Galactic high-energy cosmic rays). The ''fluctuations'' events were usually detected at the detection threshold and are absent in the data of other space experiments (primarily in the IBIS/ISGRI and SPI–ACS data) and were excluded from our analysis. A detailed study of the events associated with the interaction of the detectors with charged particles is beyond the scope of this paper.

We classified the detected events based on the following criteria: \\
(1) \textit{\textbf{The duration}}. The parameter \t90 was used as a parameter characterizing the duration of an event (Koshut et al. 1996). This parameter usually lies within the range (0.1, 100) s for GRBs and in the interval (0.01, 1) s for SGR and AXP bursts. The events associated with the interaction of the detectors with charged particles have a duration from fractions of milliseconds to fractions of a second, depending on the type. \\
(2) \textit{\textbf{The spectral hardness}}. The ratio of counts in the ranges (100, 1000) and (20, 100) keV was used as a parameter characterizing the spectral hardness. For SGR and AXP bursts, this parameter is considerably smaller than one. Since the GRB spectra are highly varied, the hardness parameter can change in a wide range and has value of about one on average. This parameter is considerably smaller than one for the first type of detector interactions with charged particles and considerably larger than one for the other two types. \\
(3) \textit{\textbf{The distribution of counts over the detectors}}. To quantitatively estimate the distribution of event counts over the detectors, we used the ratio of the maximal photon count rate in a single detector to the mean photon count rate in a single detector. For ''real'' gamma-ray events (GRBs, SGR and AXP bursts) in the SPI field of view (including those at the edge of the field of view), this parameter, as a rule, lies in the range (1–3). For events of the second type of detector interactions with charged particles, the value of this criterion usually exceeds considerably 3. \\
(4) \textit{\textbf{The pattern of spectral evolution}} was estimated at this step in most cases visually and served as an additional criterion for the selection of events associated with the interaction of the detectors with charged particles. Most of the ''real'' events are characterized by the evolution of their energy spectrum from hard to soft (see the Section ''Spectral Evolution of Gamma-Ray Bursts'' below). The spectral evolution of the events associated with the interaction of the SPI detectors with (presumably) protons, as a rule, has a directly opposite pattern. The events associated with the interaction of the SPI detectors with (presumably) Galactic high-energy cosmic rays are observed as spectral lines with energies of 55, 64, and 198 keV corresponding to the nuclear reactions of thermal neutron capture by Ge nuclei (Weidenspointner et al. 2005), where the neutrons are produced by cascade reactions. A parameter representing the ratio of the number of counts in the interval (0, 50) ms to the number of counts in the interval (-50, 0) ms in a narrow energy range, (195, 201) keV, was also used to select such events. This criterion was applied only for the selection of events on time scales of 10 and 100 ms, because these events have a duration, on average, of about 50 ms. \\
(5) \textit{\textbf{The peculiarities of the event detection rate}} were evaluated visually. Most of the SGR and AXP bursts are observed during the activity of the corresponding source, which lasts for about a month. A highly nonuniform detection rate is typical of the events associated with the interaction of the SPI detectors with protons and electron beams. The GRB detection rate, on the contrary, is constant. This is due to their cosmological nature. \\
(6) \textit{\textbf{The presence of events in the SPI-ACS and IBIS/ISGRI data}}. Most (more than 90\%) of the ''real'' gamma-ray events (confirmed by other space experiments) were also detected by the IBIS/ISGRI and about a third of the events were detected by the SPI-ACS. The presence and localization of an event in the IBIS/ISGRI data is a reliable indicator of a ''real'' gamma-ray event. Some of the events associated with the interaction of the SPI detectors with electron beams (\si 30\% of the events) and Galactic high-energy cosmic rays (\si 70\% of the events) were detected in the SPI-ACS data.

For the candidates of real gamma-ray events, we searched for their confirmations in the known catalogs of gamma-ray transients:\\
(1) the master list (Hurley 2010) that is the list of all the known GRBs from 1991 to the present day; \\
(2) the catalog of X-ray bursts of the IBIS/ISGRI experiment (Chelovekov and Grebenev 2011); \\
(3) the catalog of bursts of SGR 1806-20 (Molkov 2010) detected with IBIS/ISGRI; \\
(4) the online IBIS/ISGRI catalog of gamma-ray events (http://ibas.iasf-milano.inaf.it/); \\
(5) the online catalog of SPI–ACS triggers (http://www.isdc.unige.ch/integral/ibas/cgi-bin/ibas\_acs\_web.cgi).

For all candidate events, we also carried out an independent search for confirmations in the SPI–ACS and IBIS/ISGRI data. As has been pointed out above, the fields of view of the SPI, IBIS/ISGRI, JEM-X, and OMC aperture telescopes are aligned and SPI–ACS records photons almost from any direction. Therefore, one event can be present in the data of several telescopes at once. The IBIS/ISGRI field of view is slightly smaller in size than the SPI field of view and the sensitivity energy ranges of these detectors overlap in the range (20, 200) keV, with the efficiency of the ISGRI detector with respect to SPI in this energy range being higher. From this it follows that most of the ''real'' gamma-ray events with a high intensity in the range (20, 200) keV detected in the SPI data should also be present in the IBIS/ISGRI data. The events recorded at the edge of the SPI field of view constitute an exception. On the other hand, the effective area of the SPI–ACS detector increases with increasing angle between the source direction and the aperture telescope axis. Since the SPI–ACS sensitivity energy range (80, 10000) keV overlaps with the sensitivity energy ranges of IBIS/ISGRI (20, 200) keV and SPI (20, 8000) keV, a ''real'' event detected in the SPI data at the edge of or outside the IBIS/ISGRI and SPI fields of view should also be observed with a high probability in the SPI–ACS data. Thus, a gamma-ray event detected in the SPI data should be present with a high probability either in the IBIS/ISGRI data (if the source of the event is within the IBIS/ISGRI field of view and the event has a soft spectrum) or in the SPI-ACS data (the detection probability increases with increasing angle between the burst source direction and the aperture telescope axis). Note that very intense events are seen in both IBIS/ISGRI (provided that the source of the event is within the field of view) and SPI–ACS data. The search for confirmations in the SPI–ACS and IBIS/ISGRI data allows the candidates for real gamma-ray events detected in the SPI data to be additionally tested. Therefore, if an event was detected neither in the SPI–ACS data nor in the IBIS/ISGRI data (while it had the properties described above needed for its detection in these experiments), then it was excluded from the subsequent analysis and transferred to the first group of events (fluctuations).

When an event was detected in the light curve constructed from the IBIS/ISGRI data, it was localized using these data. We failed to localize some of the events using IBIS/ISGRI, despite their high statistical significance. As a rule, this is because the event was recorded at the edge of or outside the IBIS/ISGRI field of view. In several cases, we localized the events based on the SPI data only. Finally, we used the IPN3 triangulation data (Hurley 2011) to restrict localization of several events.

\section*{\textbf{RESULTS}}

\subsection*{\textbf{The Catalog of Confirmed Bursts and Candidates for Bursts \\from SGR 1806-20 and AXP 1E\_1547.0-5408}}

We detected 223 bursts from SGR 1806-20 (Table 1) and 23 bursts from AXP 1E\_1547.0-5408 (Table 2). Tables 1 and 2 provide the event dates, times, and durations. All of the events were confirmed and localized in the IBIS/ISGRI experiment. Some of the events were also detected in the SPI–ACS data.

Table 3 presents a list of candidates for SGR and AXP bursts. The candidates were selected in accordance with the observed properties of the confirmed bursts from SGR 1806-20 and AXP 1E\_1547.0-5408, namely the duration of individual pulses in the event is within the range (0.01–3) s, the fraction of photons with an energy above 200 keV is negligible (a soft spectrum), and the distribution over the detectors is nearly uniform. Some of the events were also identified in the IBIS/ISGRI and SPI–ACS data, but we failed to localize their sources on the celestial sphere based on the IBIS/ISGRI data. This may be because the burst sources were at the edge of the IBIS/ISGRI field of view. Some of the events were recorded in the activity period of SGR 1806-20 and AXP 1E\_1547.0-5408, when these sources were within the SPI field of view, which does not rule out the possible association of the detected events with the activity of known sources. The possible association with the activity of a known source is specified for such events in Table 3.

\subsection*{\textbf{The Catalog of Confirmed GRBs}}

We found 48 GRBs confirmed by other space experiments (Table 4). Table 4 provides the main observed properties for this group of events. Columns 1 and 2 give, respectively, the burst name and the time (UTC) of the trigger generated by the algorithm of searching for events in the SPI data. Column 3 gives the GRB duration parameter \t90. \t90 is the time it takes to record 90\% of the total counts during the entire burst (Koshut et al. 1996). Here, we calculated the burst duration mainly from the IBIS/ISGRI data in the energy range (20, 200) keV, because this detector has a much higher sensitivity in this energy range than the SPI detector, which allows the GRB duration to be determined with a higher accuracy. The SPI data were used to determine the duration of only those events that were not recorded in the IBIS/ISGRI experiment. Column 4 gives the status of the X-ray, optical, and radio GRB afterglow observations denoted by X, O, and R, respectively, in the case where the corresponding emission component is detected. Columns 5 and 6 provide the celestial equatorial coordinates RA and Dec of the burst source if it was localized. Column 7 gives the localization accuracy of the burst source. Column 8 gives the angle between the burst source direction and the center of the SPI field of view. We localized the burst source based on the IBIS/ISGRI and SPI data if the burst under study was not localized previously in other papers or the accuracy of our localization was higher than that of the known localization. We localized a total of eight GRBs based on the IBIS/ISGRI data. We failed to localize some of the events probably because the burst source was at the edge of or outside the SPI and IBIS/ISGRI fields of view. GRB 031111A and GRB 081110C were localized by the IPN3 triangulation method (for a description of the method, see, e.g., Hurley et al. 2005). Interestingly, the coordinates of the GRB 031111A source obtained by this method were outside the SPI and IBIS/ISGRI fields of view at the instant the burst was detected, implying that this burst was recorded by the SPI detector outside the field of view through SPI–ACS. This is confirmed by the high intensity of this burst in the SPI–ACS data. Apart from GRB 031111A, GRB 090902B was recorded outside the field of view. Both bursts were also detected in the IBIS/ISGRI data but were not localized. For GRB 081110C and GRB 070418B, one of the two triangulation annuli intersection regions turned out to have an intersection with the SPI field of view, with the center of the region being at an angular distance of 13\gr~and 15\gr~from the center of the field of view, respectively. The coordinates of the center of this error region and its corner coordinates are given in Table 4. Thus, GRB 081110C could be recorded at the edge of the SPI field of view but might not fall within the IBIS/ISGRI field of view, while GRB 070418B was not localized in the SPI and IBIS/ISGRI experiments, because it was at the very edge of the field of view.

Table 5 presents the statistics of observations for the GRBs investigated by other space telescopes. Columns 1 and 2 give, respectively, the GRB name and the name of the space observatory/telescope that, apart from the INTEGRAL observatory, recorded the event. Column 3 specifies whether the event has ever been localized by the IPN3 triangulation method (Hurley et al. 2005). Columns 4 and 5 specify whether the GRB is observed in the SPI–ACS or IBIS/ISGRI data, respectively. GRB 031219A, GRB 050213B, and GRB 081110C were not detected in the IBIS/ISGRI data. GRB 081110C is a short burst and has a hard spectrum. This burst was not recorded by the ISGRI detector, although it could be within the IBIS/ISGRI field of view. GRB 031219A and GRB 050213B were not localized and are probably outside the IBIS/ISGRI field of view. Column 6 specifies whether the GRB was localized in the IBIS/ISGRI experiment. Twelve bursts (out of the 45 recorded by IBIS/ISGRI) were not localized in the IBIS/ISGRI experiment, although they were observable in the light curve. These events were probably recorded at the edge of or outside the IBIS/ISGRI field of view, which makes their localization difficult. Column 7 specifies whether the burst is contained in the IBIS/ISGRI catalog of GRBs (Vianello et al. 2008) containing 56 events recorded until September 2008. We found 40 GRBs in the SPI data for this period, with 17 of then being absent in the IBIS/ISGRI catalog (Vianello et al. 2008).

Table 6 presents the results of our spectral modeling for the GRBs based on the SPI data. Column 1 gives the GRB name. Column 2 gives the photon spectral index (the exponent in the powerlaw spectral model). Band's model (Eq. (4); see also Band et al. 1993) was used to fit the spectrum of GRB 071003A; two parameters, the photon indices $\alpha$ and $\beta$, are given for this burst in column 2. Column 3 gives the position of the exponential cutoff if the power-law model with an exponential cutoff (Eq. (3)) was used or the position of the break in Band's model (Band et al. 1993). Column 4 gives the observed fluence in the energy range (20, 200) keV calculated within the spectral model used. Column 5 gives the reduced chi-square value and the corresponding number of degrees of freedom. Column 6 specifies the investigated GRB component. For three bright GRBs (GRB 041218A, GRB 050525A, and GRB 080723B), we constructed the spectra both for the entire event and for its individual, well separated (in time) components. The light curve and energy–time diagram for GRB 050525A and GRB 080723B are presented in Figs. 1 and 2, respectively. The spectra of GRB 050525A and GRB 080723B are presented in Figs. 3 and 4, respectively. To construct the spectra, we used the spi\_science\_analysis and spi\_grb\_analysis modules of the standard OSAv10.0 software package (http://isdc.unige.ch/integral/analysis\#Software) developed to process the INTEGRAL data. Our spectral modeling was performed with the XSPEC software. We constructed and investigated a total of 39 spectra for the events where the statistical significance allowed this to be done. To fit the spectra of 22 events, we used a simple power-law model with two independent parameters, A and $\alpha$ (Eq. (2)). In 17 cases, the power-law model describes unsatisfactorily the observed spectrum. As an alternative model, we used a power-law model with an exponential cutoff with three independent parameters, A, $\alpha$, and $E_{c}$ (Eq. (3)). This model describes well 16 of the 17 spectra (see Figs. 3–5). To fit the spectrum of GRB 071003A, we used Band’s more complex model (Eq. (4); see also Band et al. 1993) with four independent parameters, A, $\alpha$, $\beta$, и $E_{c}$ (Fig. 6):

\begin{eqnarray}
N(E) = AE^{-\alpha}
\label{one}.
\end{eqnarray}

\begin{eqnarray}
N(E) = AE^{-\alpha}\exp\left(-\frac{E}{E_{c}}\right)
\label{one}.
\end{eqnarray}

\begin{eqnarray}
N(E) = \left\{ \begin{array}{rl}
A\left(\frac{E}{100 keV}\right)^{\alpha}\exp\left(-\frac{E}{E_{c}}\right), &\mbox{ ($\alpha$-$\beta$)E$_{c}\geq$ E } \\
A\left(\frac{(\alpha-\beta)E_{c}}{100 keV}\right)^{\alpha-\beta}\exp(\beta-\alpha)\left( \frac{E}{100 keV}\right)^{\beta}, &\mbox{ ($\alpha$-$\beta$)E$_{c}\leq$ E }
\end{array} \right.
\label{one}.
\end{eqnarray}

Figure 7 presents the distribution of the confirmed GRBs in duration. The distribution is bimodal in pattern: there are 5 and 43 events in the groups of short and long bursts, respectively. The curve was fitted by the sum of two log-normal distributions. The centers of the log-normal distributions have the following coordinates: T$_{90}^{short}$ = 0.39$_{-0.09}^{+0.12}$ s, T$_{90}^{long}$ = 19.23$_{-1.04}^{+1.09}$ s. The minimum of the model of bimodal curve is at a point \si 1.5 s. The fraction of short GRBs in the total number of bursts is 10\%, which formally exceeds the fraction of short bursts in the IBIS/ISGRI experiment by a factor of 2.5 but is smaller than the fraction of confirmed short bursts in the SPI–ACS experiment (Minaev et al. 2009, 2010b), 16\%.

\subsection*{\textbf{The Catalog of GRB Candidates}}

Table 7 presents a list of the GRB candidates detected in the SPI/INTEGRAL experiment that contains information of 160 events. The table provides the trigger date and time (UTC), the event duration and significance in standard deviations. The candidates were selected in accordance with the observed properties of the GRBs confirmed by other space telescopes, namely the duration of individual pulses in the event is more than 5 ms, a hard energy spectrum for events with a duration of less than 2 s (this criterion also allows the events associated with the SGR activity to be rejected), and the distribution over the detectors is nearly uniform. The overwhelming majority of the selected events (94\% of all GRB candidates) have a duration of less than 2 s. Out of these events, 94 have a duration of less than 0.1 s and a low significance (on average, 6.5$\sigma$). Some of these candidates were also identified in the IBIS/ISGRI and SPI–ACS data, but we failed to localize their sources on the celestial sphere based on the IBIS/ISGRI data. Since one of the criteria for the selection of events into the group of short GRB candidates was a hard energy spectrum, the chances for successful detection of these events in the IBIS/ISGRI experiment are low. We also searched for confirmations in the Konus/WIND and GBM/Fermi experimental data. None of the corresponding events was detected in these experiments.

\subsection*{\textbf{Discussion}}

We found 48 GRBs confirmed by other space experiments in the SPI/INTEGRAL data; seven of these bursts have been detected for the first time in the SPI and/or IBIS/ISGRI data, with GRB 060221С and GRB 070912A being ''new'', previously unreported GRBs. Minaev et al. (2012) investigated GRB 070912A based on the INTEGRAL and Konus/WIND data. GRB 060221C is a typical short burst (\t90 = 0.3 s, Fig. 8). Thus, there are actually GRBs that were not identified by IBAS in the SPI data.

In Fig. 9, the fraction of short GRBs in various experiments is plotted against the lower energy threshold of the detector trigger algorithm (for more details, see Minaev et al. 2010b). It follows from Fig. 9 that the fraction of short bursts increases with increasing lower energy threshold of the detector triggers. This could be because short bursts have, on average, a harder energy spectrum. The point corresponding to the SPI experiment corresponds well to the observed dependence. The fraction of short GRBs confirmed by other space telescopes in the SPI–ACS experiment is 16\% (Minaev et al. 2009, 2010a, 2010b), which is considerably smaller than the value of about 30\% expected from the dependence. Minaev et al. (2009, 2010a, 2010b) showed that the fraction of short GRBs in the SPI–ACS experiment could reach 45\% if the candidates for short GRBs unconfirmed by other experiments were also taken into account. Minaev et al. (2010b) estimated the fraction of short bursts by taking into account the unconfirmed events to be 31\%, in good agreement with the dependence of the fraction of short bursts on the lower energy threshold of the detector triggers. An upper limit for the fraction of short GRBs is specified for the GRANAT/WATCH experiment (Sazonov et al. 1998). The fraction of short bursts in the GBM/Fermi experiment is 18\% (Paciesas et al. 2012).

We selected 160 GRB candidates; 151 of these events belong to short bursts, with 94 events having a duration of less than 0.1 s and a low significance (on average, 6.5$\sigma$). It may well be that some of the events are random background fluctuations. The total number of events detected in the SPI data is greater than the presumed number of random statistical triggers. This may imply that events unrelated to the random background variations are actually present in the data. However, it is impossible to determine which event is a fluctuation without a reliable confirmation of the nature of the event (for example, the localization of the source or the simultaneous detection by a different experiment). In this case, we can only consider the statistical probability for the presence of a certain number of fluctuation-related events in a given group.
\clearpage

\section*{\textbf{SPECTRAL EVOLUTION OF GAMMA-RAY BURSTS}}

\subsection*{\textbf{Event Selection}} \noindent

To investigate the spectral evolution, we selected 28 most intense GRBs from the catalog. Most of them have a complex, multipulse structure of their light curves. For these events, the data were processed both for the entire event and for its individual pulses if the signal level between them dropped to the background value.

We investigated the most intense bursts based on the SPI data and the remaining ones based on the IBIS/ISGRI data. Only three bursts were analyzed using both SPI and IBIS/ISGRI data. A total of 43 events were analyzed. Here and below, the specific object of investigation rather than the GRB as a whole is called an event. For example, the same GRB analyzed in both SPI and IBIS/ISGRI data comprises two events; we also call the investigation of individual episodes in multipulse bursts the investigation of various events.

\subsection*{\textbf{The Methods of Investigation}} \noindent

The conventional method of investigating the spectral evolution of GRBs consists in determining the spectral lag between the light-curve profiles in various energy channels by means of a cross-correlation analysis (Band 1997). For example, four channels were used in the BATSE experiment: (25, 50), (50, 100), (100, 300), and (300, 1000) keV (below referred to as channels 1, 2, 3, and 4, respectively). Only three quantities can be determined by this method: the lags between channels 2 and 1, 3 and 1, and 4 and 1. For most GRBs, the intensity of their emission in channel 4 is very low and it is often excluded from consideration. Thus, the shape of the dependence of the lag on energy channel cannot be established by this method. We can only get an idea of the general behavior of spectral evolution for a GRB, i.e., determine whether the spectral evolution is observed in a given event and ascertain what its pattern is — the evolution from hard to soft (positive lag) or from soft to hard.

Here, we propose a modification of the conventional method for investigating the spectral evolution of GRBs that allows the spectral evolution to be studied in more detail if there is the possibility of multichannel detection and sufficient statistics. This method was first discussed in Minaev et al. (2013).

A cross-correlation analysis of the light curves in various energy channels also underlies the proposed method. It differs from the conventional method in that a larger number of narrow energy channels are used. The channel width is chosen arbitrarily and its choice depends only on the event intensity—the width is smaller for more intense events. Up to 25 channels are used in analyzing the brightest events, implying that up to 24 spectral lags can be calculated. If the energy dependence of the spectral lag has a complex shape, then the method we proposed will allow us to detect this fact and to establish the type of this dependence. The spectral evolution in narrow spectral channels was also investigated by Preece et al. (2013), who studied GRB 130427A.

\subsection*{\textbf{Data Processing}} \noindent

The data processing consists of the following steps:

\textbf{\textit{(1) Constructing the energy–time diagram}}. The main properties of a GRB are estimated from the shape of the energy–time diagram: the intensity, the duration, the number of pulses/activity episodes, the spectral hardness, and the energy band in which the burst is observed. If the burst consists of several activity episodes (separate pulses or groups of pulses between which the signal level drops to the background value), then each such episode was considered a separate event and was investigated individually. The next processing steps are used in investigating precisely such separate events and not only in investigating the GRB as a whole.

Owing to the photon-by-photon recording of events and the high spectral resolution of the SPI and IBIS/ISGRI telescopes, we can investigate the spectral evolution of an event with high time and energy resolutions dependent only on the properties of the event itself.\\

\textbf{(2) The construction of a light curve in a wide energy channel and the identification of individual pulses in it}. As the model of a pulse, we used the FRED model (5) discussed, for example, in Norris et al. (2005) and Hakkila and Preece (2011), where A is the pulse amplitude, $t_{s}$ is the pulse start time, $\tau_{1}$ and $\tau_{2}$ are the parameters defining the pulse duration and shape. The parameters $\tau_{1}$ and $\tau_{2}$ define the pulse shape at the rise (the time interval $t_{s}$ < t < $t_{peak}$) and decay (the time interval $t_{peak}$ < t) phases, respectively. The parameter $\tau_{2}$ correlates with the pulse duration and, hence, can be used as a duration parameter:

\begin{eqnarray}
I(t) = A\lambda exp(-\frac{\tau_{1}}{t-t_{s}}-\frac{t-t_{s}}{\tau_{2}}), ~ \lambda = exp(2(\frac{\tau_{1}}{\tau_{2}})^{\frac{1}{2}}), ~ t - t_{s} > 0
\label{one}.
\end{eqnarray}

The technique for determining the number of separate pulses is as follows. Since the light curves of GRBs usually consist of pulses with various durations and intensities, we identified the pulses on various time scales. First we constructed the light curve with a high time resolution in which we identified the most significant (usually shortest) pulses that overlapped insignificantly. In some of the bursts, the pulses overlap strongly with one another and they cannot be identified even in the light curve with a high time resolution. In this case, the number of pulses was estimated visually and is a lower limit. Nonparametric methods are used in other studies of the number of pulses (see, e.g.,Mitrofanov et al. 1998; Scargle 1998).

For bursts with well-separated pulses in the case where the most intense pulse was successfully identified or, more specifically, after fitting this pulse by the model dependence (5), the procedure was repeated for the light curve obtained by subtracting the model of the successfully identified pulse from the original light curve. This procedure was repeated until there were no significant (more than five standard deviations) excesses in the light curve of the residual on time scales of the order of the bin duration. In the case where there were no significant excesses on these time scales, we grouped the bins of the light curve under consideration to find and identify longer pulses.

The number of pulses identified in the light curves of the GRBs being investigated is provided in Table 8. For bursts with a complex structure of their light curves consisting of a large number of strongly overlapping pulses that cannot be separated, we give the minimum number of pulses estimated visually (these events are marked by the letter \textbf{\textit{v}}) or the number of pulses that we managed to identify (these events are marked by \textit{\textbf{f}}). The letter \textit{\textbf{s}} on GRB 041212 means that we identified the pulses in the light curve of this burst based on the SPI-ACS data. The efficiency of the SPI and IBIS/ISGRI aperture telescopes decreases with increasing angle between the telescope axis and the burst source direction, while the SPI–ACS efficiency increases with this angle. GRB 041212 (Fig. 10) was detected at the edge of or entirely outside the SPI and IBIS/ISGRI fields of view, because our attempt to localize this burst with the SPI and IBIS/ISGRI instruments turned out to be unsuccessful. Therefore, the significance of this burst is considerably higher in the SPI–ACS experiment and we analyzed GRB 041212 based on the SPI–ACS data.

\textbf{(3) The construction of light curves} by taking into account the observed properties of an event specified in step (1) in narrow energy channels with a high time resolution.

\textbf{(4) Calculating the set of spectral lags between the generated light curves by means of a cross-correlation analysis.} For a detailed description of the application of a cross-correlation analysis to determine the spectral lag, see, e.g., Band (1997). Here, we note that the method is based on constructing the cross-correlation function (CCF) of two light curves in different energy channels. The position of the CCF peak determines the time shift between the light-curve profiles, i.e., the spectral lag.

The error in definition of the spectral lag is calculated by the Monte Carlo method. We simulate the light curves in two energy channels based on the statistical properties of the signal of the original observed light curves.

The number of counts in each bin of the simulated light curve is defined as a random variable distributed according to the Poissonian law with a mean equal to the number of counts in the bin of the original light curve. Once the light curves in two energy channels have been simulated, we perform a cross-correlation analysis of these light curves and calculate the lag between them. This procedure is repeated a thousand times (a thousand pairs of light curves are simulated) and the distribution of the derived lags is constructed. The distribution is fitted by a Gaussian whose full width at half maximum reflects the scatter of derived lags. It is this quantity that is subsequently used as the error in definition of the spectral lag.

Here, we chose the first, ''softest'' channel as the reference one relative to which the set of lags is calculated, i.e., we calculate the spectral lags between the softest channel and the remaining ones. The processing results do not depend on the choice of a reference channel.

\textbf{(5) Constructing and fitting the energy dependence of the spectral lag}. Figure 11 shows the energy dependence of the spectral lag for GRB040323 constructed from the IBIS/ISGRI data. The increase in the lag with energy (a positive slope) corresponds to a positive lag — the emission in soft energy channels lags behind the emission in hard ones. The horizontal straight line of the dependence (zero slope) means the absence of a spectral lag — the time profiles in various energy channels are not shifted in time relative to each other.

In most (37 of 43) cases, the energy dependence of the spectral lag is fitted by a logarithmic law (Eq. (6), Fig. 11) in the entire investigated energy range:

\begin{eqnarray}
lag = A\log(E)+B
\label{one}.
\end{eqnarray}

\textit{\textbf{A}} is a parameter characterizing the spectral evolution of GRBs. The sign of this parameter reflects the type of spectral evolution — a positive value means a positive spectral lag, i.e., the spectral evolution from hard to soft. Its absolute value reflects the ''speed'' of the spectral evolution — the higher it is, the more rapidly the spectrum evolves; zero means the absence of spectral evolution. Below in the text, we will call the parameter \textit{\textbf{A}} the ''spectral lag index''.

In the remaining six cases, the dependence is much better described by a logarithmic law with a break, i.e., $lag \sim A\log(E)$ in one segment of the gamma-ray spectrum and $lag \sim B\log(E)$ in the other segment, where A $\neq$ B. As the function describing this dependence, we use the sum of two logarithmic functions with an exponential cutoff (Eq. (7)), where $E_{cut}$ is the break energy and \textit{\textbf{C}} is the break sharpness. The larger the value of \textit{\textbf{C}}, the sharper the transition from one dependence to the other. In our fitting, \textit{\textbf{C}} was fixed for all events and set equal to 100. The coefficient \textit{\textbf{B2}} is not independent and is expressed in terms of the remaining model parameters via Eq. (8). Examples of the energy dependence of the spectral lag that is fitted by Eq. (7) are presented in Figs. 12-14:

\begin{eqnarray}
lag = (A_{1}\log(E)+B_{1})\exp\left[\left[-\frac{\log(E)}{\log(E_{cut})}\right]^{C}\right]+(A_{2}\log(E)+B_{2})\left[1-\exp\left[\left[-\frac{\log(E)}{\log(E_{cut})}\right]^{C}\right]\right]
\label{one}.
\end{eqnarray}
\begin{eqnarray}
B_{2} = E_{cut}(A_{1}-A_{2})+B_{1}
\label{one}.
\end{eqnarray}

Obviously, the model with a break (Eq. (7)), i.e., the model with a larger number of parameters being fitted will better describe the observed dependence for all events. In most cases, however, it is more preferable to use a simpler model with a smaller number of parameters (Eq. ( 6)). We used the statistical criteria $AIC_{c}$ and BIC (Liddle 2007). They allow one to assess whether it is appropriate to use a more complex model with a larger number of parameters. They can be calculated from Eqs. (9) and (10), where $\chi^{2}$ is the value of the functional, \textit{\textbf{k}} is the number of model parameters, and \textit{\textbf{N}} is the number of points used to fit the dependence. If the parameter $AIC_{c}$ or BIC calculated for the simple model (6) is larger than the corresponding parameter of model (7), then this means that, in this case, it is appropriate to use the more complex model (7):

\begin{eqnarray}
AIC_{c} = \chi^{2}+2k+\frac{2k(k+1)}{N-k-1}
\label{one}.
\end{eqnarray}
\begin{eqnarray}
BIC = \chi^{2}+k\ln(N)
\label{one}.
\end{eqnarray}

The processing results are presented in Table 8. Column 1 gives the GRB name; column 2 gives the name of the telescope whose data were used in our analysis; column 3 gives the number of components in the GRB light curve that was determined by identifying and fitting the pulses or was estimated visually for bursts with strongly overlapping pulses (see step 2 of the processing algorithm); column 4 specifies the component in the light curve that was used in our analysis (full means that the GRB as a whole was investigated, 1st, 2nd, etc. mean that the first, second, etc. components were investigated; the letters \textit{\textbf{g}} and \textit{\textbf{p}} mean that, respectively, several pulses and one pulse are analyzed); columns 5 and 6 provide the slopes in the fit to the spectral lag–energy relation (if the simple logarithmic model (Eq. (6)) was chosen, then the value of only one slope is given); column 7 gives the energy at which a break in the spectral lag–energy relation is observed (topical for the model with a break; Eq. (7)); column 8 gives the $\chi^{2}$ value with the indicated number of degrees of freedom for the model of the spectral lag–energy relation.

For comparison, all of the events investigated by the method described above were analyzed by the classical method — the lags between the channels (25, 50), (50, 100), (100, 300), and (300, 1000) keV (the equivalent energy channels of the LAD/BATSE detectors) were obtained. The results are presented in Table 9. Column 1 gives the GRB name (the year, month, and day of detection); column 2 gives the name of the telescope whose data were used in our analysis; column 3 specifies the component in the light curve that was used in our analysis (full means that the GRB as a whole was investigated, 1st, 2nd, etc. mean that the first, second, etc. components were investigated).

The results obtained when investigating the spectral evolution in four equivalent LAD/BATSE channels are consistent with those obtained when investigating it by the new method (see Tables 8 and 9).

\subsection*{\textit{\textbf{DISCUSSION}}} \noindent

We investigated the spectral evolution of the brightest GRBs contained in the catalog. A total of 43 events were investigated. The energy dependence of the spectral lag was constructed for each event. This dependence was fitted by two functions, a simple logarithmic one (Eq. (6)) and a logarithmic one with a break (Eq. (7)). The choice of a model in each specific case depended on the values of the statistical criteria $AIC_{c}$ and BIC (Liddle 2007). The processing results are presented in Table 8.

In 6 of the 43 cases, the energy dependence of the spectral lag is not described by a simple logarithmic law (Eq. (6)); a break is observed in the dependence (Figs. 12–14). All of the events from this groups are characterized by a complex, multipulse structure of their light curves (see Table 8 and Fig. 2). The properties of individual pulses (spectral hardness, duration, spectral lag) for the same GRB can differ (Hakkila and Preece 2011). Consider, for example, a hypothetical situation: the event being investigated consists of two pulses. The first pulse has a soft spectrum and a higher intensity in the soft energy channel. The second pulse, on the contrary, has a hard spectrum and a higher intensity in the hard energy channel. In the case of a two-pulse event, the CCF of the light curves in different energy channels will have three peaks, one main and two secondary peaks. The main, highest peak corresponds to the situation where both pulses in the light curve are superposed on each other, i.e., the main peak shows the true relative time shift (lag) of the light-curve profiles. The secondary peaks are related to the superposition of the first pulse in the first light curve on the second pulse in the second light curve and the superposition of the second pulse in the first light curve on the first pulse in the second light curve. The height of the peaks in the CCF depends on the relative intensity of the pulses in the light curve. Therefore, if we subject our hypothetical two-pulse GRB to a cross-correlation analysis, then we will detect only one secondary peak in the CCF, the peak corresponding to the superposition of the first, soft pulse in the light curve in the soft energy channel with the second, hard pulse in the light curve in the hard energy channel. As a result, we will obtain a false, negative spectral lag that does not reflect the real picture of spectral evolution, while the true spectral lag cannot be obtained in this case. Therefore, the energy dependence of the lag can become more complicated when multipulse events are investigated, which can give rise to breaks and a negative lag.

For example, GRB 080723 consists of seven separate components between which the signal level drops to its background value (Fig. 2), with each of the components, in turn, consisting of several strongly overlapping pulses. When GRB 080723 is investigated as a whole, a significant break is observed in the spectral lag–energy relation. The spectral lag–energy relation constructed for the separate components also has a complex shape. Unfortunately, the separate components cannot be investigated individually due to a strong superposition of the pulses in them on one another.

The short GRB 070707 consists of three pulses (Fig. 16), with the spectra of the first and last pulses being softest and hardest, respectively. Recall the model of a hypothetical GRB described above. It follows from our reasoning that when investigating such an event, we should obtain a false negative spectral lag, which is confirmed by our analysis of GRB 070707 (see Table 8 and Fig. 15) — we obtained a negative spectral lag index with a 2$\sigma$ significance. We also obtained a negative spectral lag in our investigation by the conventional method (the lag between the first and third channels, see Table 9).

We found the energy dependence of the spectral lag to be well fitted by a simple logarithmic law for single-pulse events. In the case of multipulse events, the energy dependence of the spectral lag can become more complicated, which can give rise to a break in the dependence (for example, GRB 031203) and even a negative spectral lag index (GRB 070707).

For eight single-pulse events (Table 8), we found a correlation between the spectral lag index and pulse duration (Fig. 17). As the duration parameter we use $\tau_{2}$, i.e., the time constant of the fall edge of the pulse (see Eq. (5) and, e.g., Norris et al. 2005). The dependence is a power law with a power law index $1.07\pm0.10$. A correlation between the spectral lag and pulse duration was found previously while investigating the spectral evolution of GRBs by a different method of analysis using the BATSE experimental data (Hakkila and Preece 2011). They used the distance between the pulse peaks in the light curves of BATSE channels 3 and 1 as the lag parameter.

Our sample of single-pulse events contains the short GRB 081226 that also follows the dependence constructed for long bursts. In addition, the each pulses of the two-pulse GRB 080414 also fall on the same dependence. Based on the results obtained, we can hypothesize that the pulse emission mechanism is the same for short and long GRBs.

If a GRB consists of several pulses, then the spectral lag index will not correlate with the burst duration when investigating the GRB as a whole, because, in contrast to the duration, the spectral lag index is not an additive quantity. Therefore, the dependence of pulse duration on spectral lag can be used to reveal GRBs with a complex, multipulse light curve in the case where the separate pulses overlap considerably and only the enveloping curve is identified in the light curve.

\section*{\textbf{CONCLUSIONS}}

Using the algorithm developed here, we searched for and classified the events in the SPI/INTEGRAL data recorded over the period 2003-2010.

We compiled a catalog of GRBs confirmed by other space experiments and outbursts from SGR 1806-20 and AXP 1E\_1547.0-5408 in 2003–2009 containing 48, 223, and 23 events, respectively.

We compiled a catalog of candidates for GRBs and SGR bursts containing 160 and 90 events, respectively.

Six new GRBs, including two previously unknown ones, were found in the SPI data.

Seven GRBs were localized for the first time with an accuracy of 2 arcmin. For GRB 081110C and GRB 070418B, the localization was made to within one IPN3 annuli intersection.

The energy spectra were constructed for 39 events using various models (Eqs. (2)–(4)), with the parameter E$_{c}$ that allows the parameter E$_{p}$ to be determined for GRB 041218 and GRB 070925.

The energy dependence of the spectral lag for GRBs and parts of GRBs consisting of a single pulse was found to be described by the logarithmic law (6).

A new parameter characterizing the spectral evolution of GRBs, the spectral lag index А (6), was introduced.

A correlation was found between the spectral lag index and pulse duration, with the dependence being the same for short and long GRBs, which may suggest a common emission mechanism. However, to confirm the latter suggestion requires investigating a larger number of short bursts.

No negative spectral lag indices were detected for single-pulse events. The negative value may be related to a difference in the properties of individual pulses in multipulse events.

%\section*{БЛАГОДАРНОСТИ}

%Авторы выражают благодарность В.М. Лозникову и А.А. Вольновой за обсуждение работы и высказанные замечания. Работа частично поддержана грантами РФФИ 12-02-01336, 13-01-92204, и програмой РАН \glqq Происхождение, строение и эволюция объектов Вселенной\grqq. Авторы выражают благодарность анонимным рецензентам за полезные замечания, способствующие улучшению статьи.

 \noindent

 \pagebreak
%****************************************************************

\clearpage
%---------------------------------------------------------------
%---------------------------------------------------------------
\begin{table}[t]

\vspace{6mm} \centering {{\bf Table 1.} List of bursts from SGR 1806-20 detected in the SPI/INTEGRAL experiment}\label{meansp}

\vspace{5mm}\begin{tabular}{c|c|c|c|c|c|c|c|c} \hline\hline
Date	&	Time	$^{*}$&	$T_{90}$	&	Date	&	Time	&	$T_{90}$	& Date	&	Time	&	$T_{90}$\\
      &   UTC   & (s)    &       &  UTC   & (s)    &       & UTC    & (s) \\\hline
2003.08.18	&	13:57:11.50	&	0.6	&	2004.08.23	&	08:56:29.69	&	0.1	&	2004.09.11	&	20:11:23.19	&	0.1	 \\
2003.08.24	&	15:30:44.23	&	0.06	&	2004.08.23	&	09:02:09.79	&	0.04	&	2004.09.15	&	00:40:05.53	&	 0.04	\\
2003.08.25	&	03:55:00.95	&	0.2	&	2004.08.23	&	09:27:49.94	&	0.04	&	2004.09.15	&	12:39:04.26	&	 0.07	\\
2003.09.20	&	13:57:54.89	&	0.1	&	2004.08.23	&	10:12:30.07	&	0.1	&	2004.09.19	&	16:38:25.92	&	 0.04	\\
2003.10.07	&	22:48:32.60	&	0.2	&	2004.08.23	&	10:38:25.09	&	0.4	&	2004.09.20	&	14:18:09.80	&	 0.03	\\
2003.10.08	&	03:27:04.09	&	0.1	&	2004.08.23	&	11:08:30.33	&	0.1	&	2004.09.22	&	20:12:16.90	&	 0.06	\\
2003.10.08	&	03:27:56.23	&	0.1	&	2004.08.23	&	11:23:10.47	&	0.2	&	2004.10.01	&	06:03:29.86	&	0.1	 \\
2003.10.08	&	20:22:06.56	&	0.1	&	2004.08.23	&	12:08:35.59	&	0.1	&	2004.10.01	&	18:45:53.22	&	0.1	 \\
2003.10.09	&	04:27:48.19	&	0.2	&	2004.08.23	&	13:21:10.16	&	0.8	&	2004.10.01	&	22:03:13.34	&	 0.05	\\
2003.10.14	&	14:31:45.45	&	0.1	&	2004.08.23	&	13:25:16.20	&	0.2	&	2004.10.02	&	10:10:27.06	&	0.1	 \\
2003.10.15	&	11:00:28.85	&	0.02	&	2004.08.23	&	13:26:03.72	&	0.2	&	2004.10.05	&	02:24:37.38	&	 0.03	\\
2003.10.15	&	12:08:26.07	&	0.2	&	2004.08.23	&	13:26:58.56	&	0.1	&	2004.10.05	&	12:57:20.14	&	0.1	 \\
2003.10.15	&	12:51:33.10	&	0.2	&	2004.08.23	&	16:17:58.00	&	0.2	&	2004.10.05	&	16:49:53.06	&	0.2	 \\
2003.10.15	&	13:18:26.73	&	0.2	&	2004.08.23	&	18:58:51.45	&	1	&	2004.10.05	&	16:49:53.06	&	0.15	 \\
2003.10.15	&	15:55:12.84	&	0.04	&	2004.08.23	&	20:48:17.71	&	0.4	&	2004.10.05	&	19:18:27.35	&	 0.4	\\
2003.10.15	&	19:23:38.58	&	0.05	&	2004.08.23	&	21:44:00.29	&	0.15	&	2004.10.05	&	19:38:57.05	&	 0.1	 \\
2004.03.09	&	10:24:08.47	&	0.2	&	2004.08.23	&	22:37:09.00	&	0.1	&	2004.10.05	&	20:04:14.03	&	 0.03	\\
2004.03.15	&	15:38:35.24	&	0.2	&	2004.08.23	&	23:01:40.34	&	0.05	&	2004.10.09	&	12:58:10.32	&	 0.1	\\
2004.03.30	&	20:37:14.45	&	0.04	&	2004.08.24	&	00:12:57.51	&	0.15	&	2004.10.14	&	06:06:35.35	&	 0.02	\\
2004.04.07	&	20:37:44.04	&	0.08	&	2004.08.24	&	02:35:22.37	&	0.05	&	2004.10.16	&	10:12:39.89	&	 0.2	\\
2004.04.07	&	21:27:11.28	&	0.2	&	2004.08.25	&	23:45:08.72	&	0.5	&	2004.10.16	&	10:51:33.98	&	 0.15	\\
2004.08.11	&	20:11:34.35	&	0.02	&	2004.08.30	&	12:03:31.32	&	0.05	&	2004.10.16	&	18:24:45.19	&	 0.3	 \\
2004.08.17	&	15:05:31.63	&	0.02	&	2004.08.30	&	14:49:37.43	&	0.1	&	2004.10.17	&	06:36:13.39	&	2	 \\
2004.08.17	&	19:35:54.27	&	0.2	&	2004.08.30	&	18:18:46.39	&	0.25	&	2004.10.17	&	06:38:49.39	&	1	 \\
2004.08.18	&	00:07:18.71	&	0.1	&	2004.08.30	&	21:46:14.73	&	0.05	&	2004.10.17	&	06:45:07.76	&	 0.03	\\
2004.08.18	&	17:10:27.17	&	0.15	&	2004.08.31	&	13:45:58.60	&	0.2	&	2004.10.19	&	16:24:30.38	&	 0.015	 \\
2004.08.18	&	20:30:10.12	&	0.2	&	2004.08.31	&	17:11:51.69	&	0.04	&	2004.10.19	&	19:02:25.94	&	 0.1	\\
2004.08.18	&	21:52:49.82	&	0.05	&	2004.08.31	&	18:32:17.89	&	0.15	&	2004.10.19	&	19:34:41.82	&	 0.05	 \\
2004.08.18	&	21:57:04.23	&	0.1	&	2004.08.31	&	20:28:57.20	&	0.2	&	2004.10.19	&	22:50:20.90	&	0.1	 \\
2004.08.18	&	22:40:26.83	&	0.05	&	2004.08.31	&	22:44:40.48	&	0.2	&	2004.10.27	&	23:11:52.48	&	 0.1	\\
2004.08.19	&	03:32:10.91	&	0.1	&	2004.09.01	&	01:21:19.81	&	0.04	&	2005.02.16	&	23:09:17.10	&	 0.1	\\
2004.08.21	&	02:37:53.36	&	0.15	&	2004.09.01	&	07:37:02.12	&	0.1	&	2005.02.17	&	00:19:56.80	&	 0.1	\\
2004.08.21	&	03:38:26.78	&	0.1	&	2004.09.01	&	17:52:21.65	&	0.15	&	2005.03.02	&	07:47:25.87	&	 0.05	\\
2004.08.22	&	15:33:08.34	&	0.15	&	2004.09.01	&	17:57:38.96	&	0.2	&	2005.03.11	&	15:40:00.83	&	 0.02	 \\
2004.08.22	&	21:32:16.68	&	0.2	&	2004.09.01	&	20:25:10.03	&	0.3	&	2005.03.20	&	11:06:04.83	&	0.3	 \\
2004.08.23	&	00:26:13.49	&	0.2	&	2004.09.02	&	07:46:46.65	&	0.2	&	2005.03.22	&	00:16:20.56	&	0.2	 \\
2004.08.23	&	01:09:46.59	&	0.1	&	2004.09.04	&	02:35:04.44	&	0.04	&	2005.03.22	&	07:23:08.24	&	 0.05	\\
2004.08.23	&	03:26:26.60	&	0.6	&	2004.09.04	&	08:42:10.45	&	0.05	&	2005.03.23	&	02:38:21.42	&	1	 \\
2004.08.23	&	05:13:15.66	&	0.15	&	2004.09.08	&	02:45:44.33	&	0.07	&	2005.03.23	&	22:49:24.28	&	 0.15	\\
2004.08.23	&	06:14:16.63	&	0.3	&	2004.09.09	&	21:18:15.38	&	0.2	&	2005.03.24	&	16:37:15.84	&	0.2	 \\\hline

\end{tabular}
\end{table}
\clearpage

\begin{table}[t]

\vspace{6mm} \centering {{\bf Table 1.} Contd.}\label{meansp}

\vspace{5mm}\begin{tabular}{c|c|c|c|c|c|c|c|c} \hline\hline
Date	&	Time	&	$T_{90}$	&	Date	&	Time	&	$T_{90}$	& Date	&	Time	&	$T_{90}$\\
      &   UTC  & (s)    &       &  UTC   & (s)    &       & UTC    & (s) \\\hline
2005.03.25	&	03:12:57.70	&	0.5	&	2006.09.16	&	23:42:56.22	&	1.1	&	2007.09.19	&	20:45:10.08	&	 1.75	\\
2005.03.28	&	03:32:35.06	&	0.1	&	2006.09.17	&	03:09:49.40	&	0.2	&	2007.09.19	&	20:45:24.99	&	0.7	 \\
2005.04.14	&	18:56:01.24	&	1.1	&	2006.09.17	&	08:16:59.21	&	0.7	&	2007.09.19	&	20:45:29.79	&	0.1	 \\
2005.04.14	&	20:17:56.25	&	0.3	&	2006.09.19	&	20:27:02.03	&	0.25	&	2007.09.19	&	20:47:30.61	&	 0.1	\\
2005.04.16	&	22:57:22.60	&	0.2	&	2006.09.19	&	20:33:21.21	&	0.4	&	2007.09.19	&	20:48:56.87	&	0.4	 \\
2005.04.16	&	22:58:50.98	&	0.1	&	2006.09.20	&	05:38:29.63	&	0.2	&	2007.09.19	&	20:51:03.69	&	 0.15	\\
2005.04.17	&	20:55:33.14	&	0.25	&	2006.09.21	&	08:39:02.87	&	0.25	&	2007.09.19	&	21:02:22.82	&	 0.1	\\
2005.04.20	&	03:44:56.25	&	0.2	&	2006.09.22	&	01:15:40.20	&	0.15	&	2007.09.19	&	21:25:25.14	&	 0.2	\\
2005.04.20	&	11:55:04.63	&	0.2	&	2006.09.23	&	02:06:16.37	&	0.35	&	2007.09.20	&	06:21:03.04	&	 0.17	\\
2005.04.21	&	05:31:37.61	&	0.7	&	2006.09.23	&	08:43:18.98	&	0.4	&	2007.09.21	&	14:52:22.68	&	0.1	 \\
2005.04.25	&	04:40:15.22	&	0.15	&	2006.09.24	&	04:48:03.55	&	0.2	&	2007.09.24	&	04:28:30.75	&	 0.2	\\
2005.04.25	&	11:03:55.84	&	0.1	&	2006.09.30	&	13:07:58.15	&	0.15	&	2007.09.24	&	07:22:56.18	&	 0.05	\\
2005.04.28	&	11:26:39.83	&	0.2	&	2006.10.04	&	08:59:13.11	&	0.15	&	2007.09.25	&	05:04:36.45	&	 0.5	\\
2005.08.11	&	13:35:48.07	&	1.5	&	2006.10.04	&	11:45:48.38	&	0.05	&	2007.09.25	&	05:10:23.49	&	 0.07	\\
2005.08.17	&	00:50:17.86	&	0.04	&	2006.11.02	&	06:38:12.16	&	0.1	&	2007.09.25	&	16:38:35.39	&	 0.2	\\
2005.08.21	&	08:31:54.59	&	0.4	&	2007.02.26	&	21:08:36.55	&	0.1	&	2007.09.26	&	11:20:34.03	&	0.1	 \\
2005.08.24	&	18:04:08.58	&	0.2	&	2007.02.27	&	07:27:48.63	&	0.1	&	2007.09.30	&	09:39:41.00	&	 0.02	\\
2005.08.28	&	12:40:19.74	&	0.1	&	2007.02.28	&	04:44:02.71	&	0.15	&	2007.10.02	&	06:31:49.62	&	 0.08	\\
2005.09.06	&	13:08:59.40	&	0.05	&	2007.03.15	&	00:54:21.74	&	0.03	&	2007.10.05	&	02:45:28.89	&	 0.1	\\
2005.09.15	&	18:42:18.36	&	0.1	&	2007.03.16	&	15:27:15.71	&	0.1	&	2007.10.29	&	08:23:36.94	&	 0.15	\\
2005.09.26	&	07:23:42.04	&	2	&	2007.03.18	&	20:29:12.99	&	0.04	&	2008.02.29	&	19:41:19.65	&	0.8	 \\
2005.10.22	&	07:25:36.62	&	0.3	&	2007.03.24	&	20:37:33.33	&	0.015	&	2008.03.19	&	15:12:23.16	&	 0.07	\\
2006.03.27	&	05:37:55.91	&	0.2	&	2007.04.03	&	19:12:13.29	&	0.15	&	2008.04.03	&	08:43:39.04	&	 0.15	\\
2006.04.22	&	15:59:12.27	&	0.3	&	2007.04.08	&	14:43:16.70	&	0.3	&		&		&		\\
2006.08.26	&	04:13:22.70	&	0.15	&	2007.08.29	&	14:03:27.83	&	0.7	&		&		&		\\
2006.08.26	&	04:25:25.46	&	1.2	&	2007.08.29	&	22:40:18.28	&	0.1	&		&		&		\\
2006.08.26	&	12:32:55.18	&	0.17	&	2007.08.29	&	23:22:21.48	&	0.1	&		&		&		\\
2006.08.26	&	18:48:46.13	&	0.4	&	2007.08.31	&	23:10:37.25	&	0.1	&		&		&		\\
2006.08.26	&	22:07:23.90	&	0.65	&	2007.09.10	&	05:31:06.64	&	0.02	&		&		&		\\
2006.08.29	&	21:57:43.66	&	2	&	2007.09.13	&	22:19:54.09	&	0.4	&		&		&		\\
2006.08.29	&	22:10:28.47	&	0.7	&	2007.09.14	&	10:46:01.32	&	0.45	&		&		&		\\
2006.08.30	&	16:37:40.60	&	0.1	&	2007.09.15	&	03:34:17.05	&	0.3	&		&		&		\\
2006.08.31	&	04:59:52.55	&	0.1	&	2007.09.15	&	13:25:35.63	&	0.03	&		&		&		\\
2006.09.04	&	15:58:43.43	&	0.2	&	2007.09.15	&	14:41:44.73	&	0.1	&		&		&		\\
2006.09.04	&	16:13:43.86	&	0.15	&	2007.09.15	&	21:34:07.41	&	0.05	&		&		&		\\
2006.09.07	&	10:36:09.99	&	0.4	&	2007.09.16	&	07:08:36.77	&	0.25	&		&		&		\\
2006.09.15	&	10:30:25.80	&	0.1	&	2007.09.16	&	09:48:03.62	&	0.35	&		&		&		\\
2006.09.15	&	17:00:01.63	&	0.3	&	2007.09.16	&	19:02:44.00	&	0.5	&		&		&		\\
2006.09.15	&	18:15:15.77	&	0.7	&	2007.09.19	&	10:01:51.91	&	0.65	&		&		&		\\
2006.09.15	&	18:58:33.19	&	0.4	&	2007.09.19	&	19:34:48.55	&	0.4	&		&		&		\\
2006.09.16	&	16:37:43.16	&	0.5	&	2007.09.19	&	19:37:29.57	&	0.3	&		&		&		\\
\hline
\multicolumn{9}{l}{}\\ [-3mm]
\multicolumn{9}{l}{$^{*}$ - The trigger time (UTС) onboard the INTEGRAL observatory}\\
\end{tabular}
\end{table}
\clearpage

\begin{table}[t]

\vspace{6mm} \centering {{\bf Table 2.} List of bursts from AXP 1E\_1547.0-5408 detected in the SPI/INTEGRAL experiment}\label{meansp}

\vspace{5mm}\begin{tabular}{c|c|c} \hline\hline
Date	&	Time	&	$T_{90}$	\\
      &  UTC   & (s)\\\hline
2009.01.25	&	04:17:16.38	&	0.1	\\
2009.01.25	&	05:22:08.79	&	0.35	\\
2009.01.25	&	05:35:08.45	&	0.02	\\
2009.01.25	&	06:38:58.75	&	0.07	\\
2009.01.25	&	10:57:27.53	&	0.04	\\
2009.01.25	&	12:55:21.61	&	0.06	\\
2009.01.25	&	17:01:54.00	&	0.01	\\
2009.01.25	&	21:52:13.37	&	0.2	\\
2009.01.29	&	16:51:40.80	&	0.05	\\
2009.01.30	&	02:34:25.28	&	0.1	\\
2009.01.30	&	06:58:16.97	&	0.15	\\
2009.01.30	&	19:29:41.68	&	0.35	\\
2009.01.30	&	19:29:43.46	&	0.05	\\
2009.02.01	&	17:10:45.54	&	0.25	\\
2009.02.01	&	21:12:56.57	&	0.35	\\
2009.02.01	&	23:13:56.80	&	0.06	\\
2009.02.05	&	19:35:40.12	&	0.04	\\
2009.02.07	&	00:54:55.80	&	0.015	\\
2009.02.07	&	20:01:20.35	&	0.03	\\
2009.02.08	&	08:09:52.38	&	0.03	\\
2009.02.08	&	11:06:28.30	&	0.02	\\
2009.03.22	&	18:56:24.25	&	0.2	\\
2009.03.22	&	22:39:16.20	&	>0.2	\\
\hline

\end{tabular}
\end{table}
\clearpage

\begin{table}[t]

\vspace{6mm} \centering {{\bf Table 3.} List of candidates for SGR bursts detected in the SPI/INTEGRAL experiment}\label{meansp}

\vspace{5mm}\begin{tabular}{c|c|c|c|c|c|c|c|c} \hline\hline
Date	&	Time	&	$T_{90}$	&	Date	&	Time	&	$T_{90}$	& Date	&	Time	&	$T_{90}$\\
      &   UTC  & (s)    &       &  UTC   & (s)    &       & UTC    & (s) \\\hline
2003.08.24$^{S}$	&	05:57:55.74	&	0.1	&	2005.12.10$^{U}$	&	03:39:50.00	&	0.02	&	2008.11.14$^{U}$	 &	 06:28:38.23	&	0.01	\\
2003.08.24$^{S}$	&	17:50:20.13	&	0.1	&	2006.04.06$^{U}$	&	23:52:54.95	&	0.01	&	2009.01.22$^{U}$	 &	 03:13:33.20	&	0.3	\\
2003.08.26$^{S}$	&	02:38:10.54	&	0.1	&	2006.04.16$^{U}$	&	01:30:39.43	&	0.015	&	2009.01.22$^{U}$	 &	 03:50:38.12	&	1.8	\\
2003.10.15$^{S}$	&	08:07:25.88	&	0.07	&	2006.04.16$^{U}$	&	09:31:07.23	&	0.02	&	 2009.01.22$^{U}$	&	 04:57:29.87	&	0.7	\\
2003.10.15$^{S}$	&	08:12:25.14	&	0.09	&	2006.04.17$^{U}$	&	16:57:03.10	&	0.02	&	 2009.01.22$^{U}$	&	 04:58:02.97	&	0.15	\\
2003.10.15$^{S}$	&	10:19:31.96	&	0.11	&	2006.04.20$^{U}$	&	10:38:56.09	&	0.02	&	 2009.01.22$^{U}$	&	 05:14:04.13	&	0.25	\\
2004.08.05$^{S}$	&	01:53:46.65	&	0.25	&	2006.04.22$^{U}$	&	01:42:23.02	&	0.02	&	 2009.01.22$^{U}$	&	 05:17:44.04	&	0.15	 \\
2004.08.09$^{U}$	&	11:31:04.31	&	0.02	&	2006.04.22$^{U}$	&	11:27:16.82	&	0.15	&	 2009.01.22$^{U}$	&	 05:17:48.24	&	0.5	 \\
2004.08.09$^{U}$	&	22:33:20.67	&	0.015	&	2006.04.22$^{U}$	&	19:08:07.83	&	0.02	&	 2009.01.22$^{U}$	&	 05:17:51.68	&	0.4	\\
2004.08.12$^{S}$	&	09:41:50.97	&	0.02	&	2006.05.05$^{U}$	&	21:00:34.93	&	0.015	&	 2009.01.22$^{U}$	&	 05:18:01.76	&	0.15	\\
2004.08.12$^{S}$	&	20:19:07.01	&	0.01	&	2006.08.29$^{S}$	&	22:10:28.47	&	0.02	&	 2009.01.22$^{U}$	&	 05:18:32.88	&	0.1	 \\
2004.08.13$^{U}$	&	15:50:48.59	&	0.01	&	2006.09.01$^{S}$	&	11:52:22.67	&	0.15	&	 2009.01.22$^{U}$	&	 05:18:39.72	&	0.9	\\
2004.08.16$^{U}$	&	14:20:53.24	&	0.01	&	2006.09.07$^{U}$	&	08:40:11.83	&	0.02	&	 2009.01.22$^{U}$	&	 05:26:52.88	&	0.2	\\
2004.08.18$^{S}$	&	08:19:50.97	&	0.02	&	2006.09.15$^{U}$	&	09:53:32.69	&	0.5	&	2009.01.22$^{U}$	 &	 06:38:27.95	&	0.25	\\
2004.08.23$^{S}$	&	05:26:47.57	&	0.015	&	2006.09.20$^{S}$	&	13:53:39.81	&	2.1	&	2009.01.22$^{U}$	 &	 06:41:02.33	&	0.9	\\
2004.08.23$^{S}$	&	17:57:00.46	&	0.02	&	2006.09.21$^{S}$	&	15:37:39.06	&	0.8	&	2009.01.22$^{U}$	 &	 06:44:36.53	&	1.4	\\
2004.08.23$^{S}$	&	18:03:09.17	&	0.2	&	2006.09.22$^{S}$	&	15:22:45.01	&	0.02	&	2009.01.22$^{U}$	 &	 06:45:12.37	&	0.15	\\
2004.08.24$^{S}$	&	12:52:39.87	&	0.02	&	2007.01.17$^{U}$	&	15:58:42.59	&	0.02	&	 2009.01.22$^{U}$	&	 06:47:57.15	&	0.25	\\
2004.09.08$^{S}$	&	21:40:52.33	&	0.3	&	2007.02.01$^{U}$	&	08:51:22.74	&	0.015	&	2009.01.22$^{U}$	 &	 06:48:15.19	&	0.45	\\
2004.09.19$^{S}$	&	07:35:55.97	&	0.025	&	2007.02.27$^{S}$	&	19:28:26.79	&	0.45	&	 2009.01.22$^{U}$	&	 06:49:49.05	&	0.15	\\
2004.10.19$^{S}$	&	17:26:27.41	&	0.15	&	2007.02.28$^{S}$	&	01:50:24.96	&	0.02	&	 2009.01.22$^{U}$	&	 07:05:56.85	&	0.25	\\
2005.03.22$^{S}$	&	03:06:40.15	&	0.1	&	2007.02.28$^{S}$	&	17:52:50.61	&	0.015	&	2009.01.22$^{U}$	 &	 07:49:40.87	&	0.4	\\
2005.04.21$^{S}$	&	09:39:57.54	&	0.1	&	2007.03.02$^{U}$	&	15:39:01.13	&	0.015	&	2009.01.22$^{U}$	 &	 08:17:29.93	&	5.5	\\
2005.08.24$^{S}$	&	17:05:58.00	&	0.7	&	2007.03.07$^{S}$	&	15:43:04.94	&	0.45	&	2009.01.25$^{A}$	 &	 00:12:48.92	&	0.1	\\
2005.09.13$^{S}$	&	05:57:51.40	&	0.5	&	2007.03.16$^{S}$	&	01:23:07.22	&	0.02	&	2009.01.25$^{A}$	 &	 05:25:05.50	&	0.025	\\
2005.09.26$^{S}$	&	21:25:43.88	&	0.1	&	2007.03.24$^{S}$	&	22:01:46.56	&	0.03	&	2009.01.25$^{A}$	 &	 23:00:36.49	&	0.07	\\
2005.10.22$^{S}$	&	02:40:56.68 &	0.1	&	2007.04.04$^{S}$	&	18:42:14.41	&	0.025	&	2009.01.29$^{A}$	 &	 22:27:27.08	&	0.07	\\
2005.10.22$^{S}$	&	23:10:44.11	&	0.01	&	2007.09.18$^{S}$	&	17:08:49.16	&	2.7	&	2009.02.21$^{U}$	 &	 15:27:34.76	&	0.1	\\
2005.11.16$^{U}$	&	02:38:19.26	&	0.02	&	2007.09.19$^{S}$	&	20:45:24.99	&	0.7	&	2009.02.23$^{U}$	 &	 03:04:41.26	&	0.15	\\
2005.12.09$^{U}$	&	01:39:46.14	&	0.03	&	2007.09.21$^{S}$	&	07:12:01.97	&	0.45	&	 2009.02.27$^{U}$	&	 13:43:35.37	&	0.35	\\
\hline
\multicolumn{9}{l}{}\\ [-3mm]
\multicolumn{9}{l}{$^{S}$ - A candidate for a burst from SGR 1806-20,}\\
\multicolumn{9}{l}{$^{A}$ - a candidate for a burst from AXP 1E\_1547.0-5408,}\\
\multicolumn{9}{l}{$^{U}$ - a candidate for a burst from an unknown source.}\\

\end{tabular}
\end{table}
\clearpage

\begin{table}[t]

\vspace{6mm} \centering {{\bf Table 4.} Observed properties of the GRBs detected in the SPI/INTEGRAL experiment}\label{meansp}

\vspace{5mm}\begin{tabular}{c|c|c|c|c|c|c|c} \hline\hline
GRB	&	Time	&	$T_{90}$			&	Afterglow${^1}$ &	RA	&	DEC	&	Radius	&	Offset	\\	
	&	UTC	    &	(s)			   &		&	(deg)	&	(deg)	&	(arcmin)	&	(deg)		\\\hline
																		
030227A	&	08:42:06.08	&	33.33	 $\pm$	2.75	&	X O	&	74.387711	&	20.484694	&	0.0005$^{8}$ 	&	8.6	\\	
031111A${^2}$	&	16:45:20.96	&	8.2	 $\pm$	4.4	&	-	&	72.43	&	17.46	&	120$^{28}$	&	53.5	\\	
031203B	&	22:01:28.96	&	34.21	 $\pm$	3.51	&	X O R	&	120.62579	&	-39.85112	&	0.012$^{9}$ 	&	11.3	 \\	
031219A${^3}$	&	04:14:42.74	&	7.8	 $\pm$	5.1	&	-	&	-	&	-	&	-	&		\\	
040223A	&	13:28:13.59	&	235	 $\pm$	3	&	X	&	249.87571	&	-41.93325	&	0.025$^{10}$ 	&	9.0	\\	
040323A	&	13:03:05.99	&	29	 $\pm$	1.93	&	-	&	208.469	&	-52.354	&	1.26$^{11}$ 	&	11.7	\\	
040421A	&	02:30:27.70	&	8.7	 $\pm$	4.6	&	-	&	-	&	-	&	-	&		\\	
040422A	&	06:58:04.70	&	9.09	 $\pm$	2.73	&	-	&	280.505	&	1.981	&	0.95$^{11}$ 	&	9.4	\\	
040730A	&	02:12:20.95	&	56.25	 $\pm$	4.45	&	-	&	238.302	&	-56.47	&	1.35$^{11}$	&	5.7	\\	
040827A	&	11:51:06.19	&	40.25	 $\pm$	3.51	&	X O	&	229.25558	&	-16.14142	&	0.003$^{12}$	&	12.1	 \\	
041211D${^2}$	&	23:57:42.26	&	6.2	 $\pm$	1.5	&	-	&	353.9	&	23.0	&	- $^{29}$	&	28.5	\\	
041212A	&	18:34:24.71	&	40.5	 $\pm$	4	&	-	&	-	&	-	&	-	&		\\	
041218A	&	15:45:59.17	&	41.4	 $\pm$	1.2	&	O	&	24.78167	&	71.34167	&	0.025$^{13}$	&	13.5	\\	
041226A	&	17:22:29.10	&	14.6	 $\pm$	8.4	&	-	&	-	&	-	&	-	&		\\	
050213B	&	19:34:46.05	&	11.1	 $\pm$	9.4	&	-	&	-	&	-	&	-	&		\\	
050502A	&	02:14:13.52	&	10.9	 $\pm$	0.16	&	O	&	202.44292	&	42.674362	&	<0.62$^{14}$ 	&	4.0	\\	
050504A	&	08:00:57.79	&	76	 $\pm$	5.2	&	X	&	201.0054	&	40.70333	&	0.11$^{15}$ 	&	4.1	\\	
050520A	&	00:06:38.93	&	52.95	 $\pm$	0.15	&	X	&	192.52583	&	30.450556	&	0.08$^{16}$ 	&	4.3	\\	
050525A	&	00:02:58.32	&	8.9	 $\pm$	0.3	&	X O R	&	278.13571	&	26.339582	&	0.002$^{17}$ 	&	14.9	\\	
051105A	&	11:05:43.05	&	24.5	 $\pm$	4.5	&	-	&	9.468	&	-40.479	&	1.1$^{11}$	&	2.8	\\	
051220A${^3}$	&	13:04:15.36	&	54.9	 $\pm$	2.3	&	-	&	350.622	&	70.116	&	2	&	11.0	\\	
060221C${^4}$	&	15:40:33.36	&	0.29	 $\pm$	0.04	&	-	&	255.053	&	-24.214	&	2	&	11.3	\\	
060306C${^3}$	&	15:22:38.94	&	0.9	 $\pm$	0.2	&	-	&	-	&	-	&	-	&		\\	
060428С	&	02:30:43.72	&	12.7	 $\pm$	1.1	&	-	&	285.227	&	-9.556	&	0.72$^{11}$	&	8.6	\\	
060901A	&	18:43:57.68	&	16.3	 $\pm$	2.1	&	X	&	287.15805	&	-6.639444	&	0.07$^{18}$ 	&	14.1	\\	
061025A	&	18:36:03.22	&	14	 $\pm$	0.8	&	X O	&	300.91177	&	-48.242981	&	0.008$^{19}$ 	&	8.5	\\	
061122A	&	07:56:53.88	&	14.3	 $\pm$	0.5	&	X O	&	303.83267	&	15.517361	&	0.008$^{20}$ 	&	7.4	\\	
070311A	&	01:52:44.25	&	35	 $\pm$	2	&	X O	&	87.53421	&	3.37508	&	0.02$^{21}$ 	&	12.2	\\	
070418B${^7}$	&	17:16:23.72	&	21	 $\pm$	8	&	-	&	302.13	&	7.37	&	- $^{7}$	& 15.0		\\	
070707B	&	16:08:38:63	&	0.7	 $\pm$	0.09	&	X O	&	267.74396	&	-68.924225	&	0.008$^{22}$ 	&	12.0	 \\	
070912A${^3}$	&	07:32:23.42	&	62.5	 $\pm$	2.3	&	X	&	264.608	&	-28.706	&	2	&	3.6	\\	
070925C	&	15:52:47.11	&	26.6	 $\pm$	4	&	X	&	253.21699	&	-22.028561	&	0.1$^{23}$ 	&	9.1	\\	
071003A	&	07:40:56.20	&	19.15	 $\pm$	1.55	&	X O R	&	301.853	&	10.948	&	2	&	13.5	\\	
071108B	&	21:41:20.40	&	7.2	 $\pm$	4.2	&	-	&	-	&	-	&	-	&		\\	
080223A${^3}$	&	19:39:29.20	&	20.85	 $\pm$	5.25	&	-	&	265.137	&	-12.827	&	2	&	15.2	\\	
080303C	&	21:34:51.51	&	23.5	 $\pm$	7.6	&	-	&	-	&	-	&	-	&		\\	
080413A${^5}$	&	02:54:20.39	&	4.3	 $\pm$	3.1	&	X O	&	287.2978	&	-27.6779	&	0.08$^{24}$	&	14.9	 \\	
080414B	&	22:33:26.51	&	10.64	 $\pm$	0.82	&	-	&	272.133	&	-18.829	&	2$^{11}$	&	11.5	\\	
080613A	&	09:35:40.72	&	27	 $\pm$	2.01	&	X O	&	213.2709	&	5.1732	&	0.01$^{25}$ 	&	3.9	\\	 \hline

\end{tabular}
\end{table}
\clearpage

\begin{table}[t]

\vspace{6mm} \centering {{\bf Table 4.} Contd.}\label{meansp}

\vspace{5mm}\begin{tabular}{c|c|c|c|c|c|c|c} \hline\hline
GRB	&	Time	&	$T_{90}$			&	Afterglow	&	RA	&	DEC	&	Radius	&	Offset	\\	
	&	UTC	    &	(s)			   &		&	(deg)	&	(deg)	&	(arcmin)	&	(deg)		\\\hline

080723B	&	13:22:34.10	&	82.77	 $\pm$	0.67	&	X	&	176.833	&	-60.245	&	1.5$^{11}$	&	7.9	\\	
081003C	&	20:48:14.55	&	23.9	 $\pm$	1.85	&	X	&	285.026	&	16.691	&	2	&	11.2	\\	
081016A	&	06:51:55.47	&	33.33	 $\pm$	1.56	&	X	&	255.5724	&	-23.3301	&	1.5	&	4.9	\\	
081110C${^3}$	&	16:18:50.80	&	0.15	 $\pm$	0.1	&	-	&	289.1	&	35.0	&	- $^{6}$	&	13.4	\\	
081226B	&	12:13:11.20	&	0.27	 $\pm$	0.05	&	-	&	25.495	&	-47.439	&	2.5	&	7.6	\\	
090107B	&	16:20:49.09	&	20.2	 $\pm$	2.11	&	X	&	284.8075	&	59.5924	&	1.5$^{26}$	&	6.7	\\	
090625B	&	13:26:25.73	&	8.4	 $\pm$	0.6	&	X	&	2.2626	&	-65.7816	&	1.7$^{27}$	&	11.4	\\	
090820A	&	00:38:47.51	&	14.3	 $\pm$	3.8	&	-	&	-	&	-	&	-	&		\\	
090902B${^2}$	&	11:05:17.40	&	22.1	 $\pm$	7.4	&	X O R	&	264.93859	&	27.32448	&	0.035$^{28}$	&	 38.1	\\	 \hline

\multicolumn{8}{l}{}\\ [-3mm]
\multicolumn{8}{l}{$^{1}$ - X, O, and R stand for X-ray, optical, and radio afterglows, respectively,}\\
\multicolumn{8}{l}{$^{2}$ - registered outside the SPI and IBIS/ISGRI fields of view,}\\
\multicolumn{8}{l}{$^{3}$ - the GRB also recorded by other space observatories, first detected in }\\
\multicolumn{8}{l}{the SPI and/or IBIS/ISGRI data,}\\
\multicolumn{8}{l}{$^{4}$ - a new, previously unknown GRB detected in the SPI data,}\\
\multicolumn{8}{l}{$^{5}$ - registered at the edge of the SPI field of view,}\\
\multicolumn{8}{l}{$^{6}$ - the burst source was localized by triangulation using the data from Hurley (2011),}\\
\multicolumn{8}{l}{the corner coordinates of the error region (RA, Dec): }\\
\multicolumn{8}{l}{(290.2, 32.1), (287.9, 37.8), (290.3, 32.0), (288.0, 37.8),}\\
\multicolumn{8}{l}{$^{7}$ - the burst source was localized by triangulation using the data from Hurley (2011),}\\
\multicolumn{8}{l}{the corner coordinates of the error region (RA, Dec): }\\
\multicolumn{8}{l}{(302.16, 6.95), (302.10, 7.78), (302.31, 7.31), (301.95, 7.43),}\\
\multicolumn{8}{l}{  }\\
\multicolumn{8}{l}{The burst source was localized in:: $^{8}$ - GCN 1907, $^{9}$ - GCN 2490, $^{10}$ - GCN 2547, }\\
\multicolumn{8}{l}{$^{11}$ - (Vianello et al. 2008), $^{12}$ - (De Luca et al. 2005), $^{13}$ - GCN 2861, $^{14}$ - GCN 3322, }\\
\multicolumn{8}{l}{$^{15}$ - GCN 3359, $^{16}$ - GCN 3434, $^{17}$ - GCN 3493, $^{18}$ - GCN 5496, $^{19}$ - GCN 5754, $^{20}$ - GCN 5849,}\\
\multicolumn{8}{l}{$^{21}$ - GCN 6190, $^{22}$ - GCN 6612, $^{23}$ - GCN 6826, $^{24}$ - GCN 7594, $^{25}$ - GCN 7872, $^{26}$ - GCN 8786,}\\
\multicolumn{8}{l}{$^{27}$ - GCN 9572, $^{28}$ - GCN 9871, $^{29}$ - Орли и др., 2011}\\
\end{tabular}
\end{table}
%---------------------------------------------------------------

\begin{table}[t]

\vspace{6mm} \centering {{\bf Table 5.} Statistics of observations for the GRBs detected in the SPI/INTEGRAL and other space experiments}\label{meansp}

\vspace{5mm}\begin{tabular}{c|c|c|c|c|c|c} \hline\hline
GRB	&	Confirmation by other 	&	IPN	&	ACS	&	ISGRI	&	ISGRI	&	Vianello et al. 2008	\\
    &   space observatories           &     &     &       & local.& ISGRI catalog \\\hline
030227A	&	RHE, Uly (RI)$^{1}$	&	-	&	-	&	+	&	+	&	+	\\
031111A	&	Kon, Uly, HET	&	+	&	+	&	+	&	-	&	-	\\
031203B	&	Kon (RI)	&	-	&	+	&	+	&	+	&	+	\\
031219A	&	Kon (RI)	&	-	&	+	&	-	&	-	&	-	\\
040223A	&	-	&	-	&	-	&	+	&	+	&	+	\\
040323A	&	Kon (RI)	&	-	&	+	&	+	&	+	&	+	\\
040421A	&	Kon, RHE (RI)	&	+	&	+	&	+	&	-	&	-	\\
040422A	&	Kon, MO	&	-	&	-	&	+	&	+	&	+	\\
040730A	&	-	&	-	&	-	&	+	&	+	&	+	\\
040827A	&	-	&	-	&	-	&	+	&	+	&	+	\\
041211D	&	MO, Kon, HET, RHE, Swi	&	+	&	+	&	+	&	-	&	-	\\
041212A	&	MO, RXTE	&	-	&	+	&	+	&	-	&	-	\\
041218A	&	RHE, Kon	&	-	&	+	&	+	&	+	&	+	\\
041226A	&	Kon, MO	&	+	&	+	&	+	&	-	&	-	\\
050213B	&	HET (RI)	&	-	&	+	&	-	&	-	&	-	\\
050502A	&	-	&	-	&	-	&	+	&	+	&	+	\\
050504A	&	-	&	-	&	-	&	+	&	+	&	+	\\
050520A	&	-	&	-	&	-	&	+	&	+	&	+	\\
050525A	&	-	&	-	&	+	&	+	&	+	&	+	\\
051105A	&	-	&	-	&	+	&	+	&	+	&	+	\\
051220A	&	MO, Kon, RHE	&	+	&	+	&	+	&	+	&	-	\\
060221C	&	-	&	-	&	+	&	+	&	+	&	-	\\
060306C	&	Swi, Kon, RHE	&	+	&	+	&	+	&	-	&	-	\\
060428С	&	Kon, RHE	&	+	&	+	&	+	&	+	&	+	\\
060901A	&	Uly, MO, Kon, Suz	&	+	&	+	&	+	&	+	&	+	\\
061025A	&	Kon, Suz	&	-	&	-	&	+	&	+	&	+	\\
061122A	&	Swi, Kon	&	+	&	+	&	+	&	+	&	+	\\
070311A	&	Kon (RI)	&	-	&	-	&	+	&	+	&	+	\\
070418B	&	Suz, Mes, Kon, MO	&	+	&	+	&	+	&	-	&	-	\\
070707B	&	MO, Kon, RHE, Swi, Mes, Suz	&	+	&	+	&	+	&	+	&	+	\\
070912A	&	Kon (RI)	&	-	&	-	&	+	&	+	&	-	\\
070925C	&	Kon, RHE, Mes	&	-	&	+	&	+	&	+	&	+	\\
071003A	&	Kon, MO, Swi, Suz	&	+	&	+	&	+	&	+	&	-	\\\hline

\end{tabular}
\end{table}

\begin{table}[t]

\vspace{6mm} \centering {{\bf Table 5.} Contd.}\label{meansp}

\vspace{5mm}\begin{tabular}{c|c|c|c|c|c|c} \hline\hline
GRB	&	Confirmation by other 	&	IPN	&	ACS	&	ISGRI	&	ISGRI	&	Vianello et al. 2008	\\
    &   space observatories           &     &     &       & local.& ISGRI catalog \\\hline
071108B	&	Kon, RHE, Swi, Suz, Agi	&	+	&	+	&	+	&	-	&	-	\\
080223A	&	MO, Kon, Mes, Suz 	&	+	&	+	&	+	&	+	&	-	\\
080303C	&	MO, Kon, Swi, Suz, Agi 	&	+	&	+	&	+	&	-	&	-	\\
080413A	&	Kon, Swi, Mes, Suz	&	+	&	+	&	+	&	-	&	-	\\
080414B	&	-	&	-	&	-	&	+	&	+	&	+	\\
080613A	&	Kon (RI)	&	-	&	-	&	+	&	+	&	+	\\
080723B	&	Kon, Mes, Agi	&	+	&	+	&	+	&	+	&	+	\\
081003C	&	Kon, RHE	&	-	&	+	&	+	&	+	&	-	\\
081016A	&	Kon, Suz	&	-	&	+	&	+	&	+	&	-	\\
081110C	&	Swi, Kon, Mes, Suz	&	+	&	+	&	-	&	-	&	-	\\
081226B	&	Swi, Suz, Fer	&	+	&	+	&	+	&	+	&	-	\\
090107B	&	Suz, Fer, Kon	&	+	&	-	&	+	&	+	&	-	\\
090625B	&	Suz, Fer, Kon	&	+	&	-	&	+	&	+	&	-	\\
090820A	&	RHE, Kon, Swi, Suz, Agi, Fer 	&	-	&	+	&	+	&	-	&	-	\\
090902B	&	RHE, Swi, Suz, LAT	&	+	&	+	&	+	&	-	&	-	\\\hline
\multicolumn{7}{l}{}\\ [-3mm]
\multicolumn{7}{l}{$^{1}$ - Background level enhancement in the detectors probably associated with a GRB.   }\\
\multicolumn{7}{l}{}\\
\multicolumn{7}{l}{The abbreviations used are: RHE for RHESSI, Uly for Ulysses, Kon for Konus,}\\
\multicolumn{7}{l}{HET for HETE-2, Swi for Swift, MO for Mars Observer, Suz for Suzaku,}\\
\multicolumn{7}{l}{Mes for Messenger, Agi for Agile, LAT for Fermi/LAT, and Fer for Fermi/GBM.}\\

\end{tabular}
\end{table}

\begin{table}[t]

\vspace{6mm} \centering {{\bf Table 6.} Spectral modeling results for the confirmed GRBs}\label{meansp}

\vspace{5mm}\begin{tabular}{c|c|c|c|c|c} \hline\hline
GRB	        &	Photon	        &	$E_{c}$	            &	Fluence (20-200 кэВ)	&	$\chi^{2}_{red}$ / dof	&	Component	    \\
	        &		   Index                 &	keV           	            &	$10^{-6}$ erg/cm$^{2}$	&		                    &		        \\\hline
030227A     &	$2.05_{-0.34}^{+0.42}$  &	                            &	$1.0_{-0.4}^{+0.2}$                  &   1.60/11	                &		full   \\
031203B     &	$1.38_{-0.21}^{+0.24}$  &                               & 	$1.7_{-0.5}^{+0.3}$                  &	0.85/13	                &		full \\
040223A     &   $2.55_{-0.76}^{+1.28}$	&	                            &	$0.4_{-0.3}^{+0.3}$                 &   1.15/8	                &     intense part \\
040323A	    &   $1.32_{-0.15}^{+0.16}$  &	  	                        &   $1.9_{-0.5}^{+0.3}$	        &   0.66/13		    	    &  full \\
040422A	    &   $1.84_{-0.22}^{+0.23}$	&	                            &   $0.95_{-0.11}^{+0.05}$              	 &   1.01/13	                &  full    \\
040730A     &	$1.16_{-0.32}^{+0.36}$  &  	                            &   $1.2_{-0.6}^{+0.4}$             	 &   0.61/11	                &  full \\
040827A	    &   $1.87_{-0.41}^{+0.50}$	&	                            &   $1.1_{-0.7}^{+0.4}$  	            &   0.96/8	                &   full \\
041218A     &	$0.73_{-0.48}^{+0.42}$	&  $75.8_{-25.5}^{+52.4}$      	&   $5.9_{-0.3}^{+0.1}$               &   1.5/21	                &   full    \\
041218A     &	$1.07_{-0.68}^{+0.54}$	&  $138.4_{-35.6}^{+72.4}$      &	$2.0_{-0.7}^{+0.3}$               & 	1.09/18	                &     1st \\
041218A	    &   $1.59_{-0.16}^{+0.17}$  &	                            &   $2.1_{-0.7}^{+0.4}$               &  	0.5/12	                &      2nd \\
041218A	    &   $1.77_{-0.24}^{+0.27}$	&                               &   $1.9_{-0.6}^{+0.3}$                 &  	1.4/10	                &     3rd     \\
050502A     &	$1.54_{-0.19}^{+0.21}$  &  	                            &   $1.3_{-0.3}^{+0.1}$      	        &   0.88/12	                &     full      \\
050504A     &	$1.33_{-0.31}^{+0.35}$	&	                            &   $1.2_{-0.4}^{+0.2}$                 & 	1.32/5	                &      full    \\
050520A	    &   $1.37_{-0.16}^{+0.18}$	&	                            &   $2.2_{-0.3}^{+0.1}$                	 &   1.08/13	                &  full    \\
050525A     &	$1.58_{-0.19}^{+0.18}$  &	$166.6_{-44.8}^{+83.2}$     &	$16.5_{-0.6}^{+0.4}$                   &	0.95/31	                &		full     \\
050525A     &	$1.25_{-0.34}^{+0.32}$  &	$123.9_{-40.9}^{+94.5}$     &	$6.3_{-0.7}^{+0.4}$	                &   0.98/28	                &		1st      \\
050525A	    &   $1.66_{-0.32}^{+0.30}$  &   $145.0_{-55.4}^{+177.9}$    &	$9.5_{-0.6}^{+0.4}$                   &	0.83/26	                &	2nd       \\
060428C     &	$0.74_{-0.61}^{+0.51}$  &  $84.83_{-35.21}^{+97.14}$	&   $2.2_{-0.8}^{+0.5}$                 & 	0.97/22	                &     full    \\
060901A	    &   $1.07_{-0.32}^{+0.27}$	&  $244.4_{-106.1}^{+135.2}$    &   $6.7_{-0.9}^{+0.6}$                &  	0.93/17	               	&  full \\
061025A	    &   $1.51_{-0.23}^{+0.27}$  &	                            &   $0.9_{-0.3}^{+0.2}$                & 	1.02/12	                &   full \\
061122A	    &	$1.07_{-0.10}^{+0.10}$	&	$188.0_{-31.4}^{+42.3}$	    &	$14.8_{-0.4}^{+0.2}$	                &	1.15/42	                &	full	    \\
070311A	    &   $1.67_{-0.29}^{+0.33}$	&	                            &   $1.7_{-0.5}^{+0.3}$                 &  	0.64/10	                &   full    \\
070418B	    &   $2.23_{-0.29}^{+0.31}$  &		                        &   $46.1_{-1.4}^{+1.0}$                 & 	0.70/18	                &  	full   \\
070707B	    &   $1.37_{-0.29}^{+0.32}$	&	                            &   $0.3_{-0.2}^{+0.1}$               	 &   0.81/13	                &    full    \\
070912A	    &   $1.36_{-0.13}^{+0.14}$	&	                            &   $1.8_{-0.2}^{+0.1}$                 &   1.35/12	                &   	full \\
070925C	    &   $0.63_{-0.58}^{+0.45}$  &	$98.1_{-42.2}^{+98.1}$	    &   $2.9_{-0.8}^{+0.5}$             	 &   0.78/9	                &  full    \\
071003A	    &   $\alpha=-0.99_{-0.10}^{+0.13}$	& 	$775.0_{-287.7}^{+707.9}$ &	  $10.9_{-0.4}^{+0.2}$           &   0.92/26	                & full    \\
            &   $\beta=-2.34_{-0.84}^{+0.53}$ &  &   &   &   \\						
080223A	    &   $2.30_{-0.44}^{+0.52}$  &		                        &   $2.1_{-0.7}^{+0.3}$                 &  	0.77/9	                & full    \\
080413A	    &   $2.06_{-0.52}^{+0.67}$  &                               &   $1.9_{-0.8}^{+0.5}$                 &   1.46/7	                & full    \\
080414B	    &   $2.55_{-0.74}^{+1.48}$	&	                            & 	$0.3_{-0.3}^{+0.4}$                 &   1.72/11	                &	full \\
080723B	    &   $1.03_{-0.07}^{+0.06}$  &   $308.6_{-47.3}^{+62.6}$     &   $26.2_{-0.6}^{+0.3}$                   &   1.12/42                 &		full     \\
080723B     &	$1.11_{-0.34}^{+0.21}$  &	$290.9_{-149.4}^{+173.2}$   &	$2.2_{-0.5}^{+0.3}$                   &	1.40/27	                &		1st      \\
080723B 	&   $0.84_{-0.14}^{+0.14}$  &  	$175.1_{-34.2}^{+43.0}$     &  	$8.4_{-0.4}^{+0.2}$                   &	1.74/37	                &		2nd       \\
080723B     &	$0.97_{-0.22}^{+0.21}$  &	$224.8_{-77.2}^{+183.4}$    &	$4.1_{-0.5}^{+0.2}$                   &	1.42/27	                &		3rd \\
080723B	    &   $0.78_{-0.21}^{+0.19}$  &   $302.7_{-103.7}^{+167.2}$   &	$3.5_{-0.5}^{+0.3}$                   &	1.52/27	                &		4th  \\
080723B     &	$1.18_{-0.18}^{+0.15}$  &	$315.9_{-114.6}^{+193.4}$   &	$5.2_{-0.6}^{+0.3}$                   &	1.18/27	                &		5th  \\
080723B	    &   $0.89_{-0.09}^{+0.07}$  &	$442.7_{-93.8}^{+184.1}$    &	$7.6_{-0.4}^{+0.2}$                   &	1.11/37	                &		6th    \\
081003C	    &   $1.61_{-0.16}^{+0.18}$  &                               &   $2.1_{-0.4}^{+0.2}$                 &	1.53/13	                & full   \\
090107A	    &   $1.67_{-0.29}^{+0.32}$  &                               &	$1.6_{-0.5}^{+0.3}$                 &   0.54/10	                &  full  \\\hline

\end{tabular}
\end{table}

\begin{table}[t]

\vspace{6mm} \centering {{\bf Table 7.} List of the GRB candidates detected in the SPI/INTEGRAL experiment}\label{meansp}

\vspace{5mm}\begin{tabular}{c|c|c|c|c|c|c|c} \hline\hline
Date	&	Time	&	$T_{90}$	      &	Significance	&	Date	&	Time	&	$T_{90}$	      &	Significance\\
        &  UTC      & (s)              &       &       &   UTC  & (s)    &      \\\hline
2003.02.17	&	02:29:03.90	&	0.2	&	5.6	&	2005.04.11	&	02:20:41.03	&	0.15	&	5.1	\\
2003.02.19	&	19:14:37.97	&	0.005	&	6.5	&	2005.04.16	&	09:50:45.47	&	0.1	&	5.1	\\
2003.02.23	&	22:14:14.83	&	0.01	&	6.1	&	2005.04.22	&	05:04:01.58	&	0.1	&	5.2	\\
2003.02.25	&	05:29:36.21	&	1.3	&	5.1	&	2005.07.18	&	05:37:38.47	&	0.1	&	5.4	\\
2003.05.30	&	17:35:10.80	&	0.01	&	6.1	&	2005.07.19	&	15:50:56.85	&	0.1	&	5.3	\\
2003.06.25	&	09:24:04.58	&	0.02	&	6.6	&	2005.09.13	&	19:23:22.29	&	12	&	8	\\
2003.07.14	&	01:46:37.29	&	0.01	&	6.2	&	2005.10.03	&	18:33:12.73	&	0.005	&	6.4	\\
2003.07.15	&	17:21:33.13	&	0.15	&	5.1	&	2005.10.15	&	12:33:59.98	&	7.5	&	5	\\
2003.07.16	&	12:45:40.85	&	0.15	&	5.3	&	2005.12.13	&	08:33:57.54	&	9	&	4.9	\\
2003.08.01	&	09:42:34.91	&	0.01	&	6.6	&	2006.02.12	&	14:30:27.22	&	0.15	&	5	\\
2003.08.11	&	17:37:21.05	&	0.015	&	7.4	&	2006.02.27	&	08:31:43.70	&	0.1	&	5.5	\\
2003.08.18	&	13:57:11.50	&	0.8	&	7.7	&	2006.03.03	&	12:09:31.41	&	0.2	&	5	\\
2003.08.31	&	19:56:25.40	&	0.01	&	6.4	&	2006.03.14	&	10:51:17.24	&	0.15	&	5.3	\\
2003.10.11	&	16:07:10.19	&	0.15	&	5	&	2006.03.20	&	09:33:09.18	&	0.15	&	5.2	\\
2003.10.15	&	17:13:51.90	&	0.001	&	7.7	&	2006.04.04	&	00:26:33.58	&	0.25	&	5.2	\\
2003.10.24	&	21:57:27.78	&	0.01	&	6.5	&	2006.04.06	&	18:39:16.26	&	0.01	&	6.6	\\
2003.10.24	&	23:41:38.81	&	0.01	&	6.5	&	2006.04.20	&	01:28:26.83	&	0.1	&	5.3	\\
2003.11.11	&	12:38:29.32	&	0.01	&	6.2	&	2006.05.16	&	08:03:04.28	&	9	&	4.1	\\
2003.12.02	&	11:14:24.17	&	0.02	&	6.8	&	2006.09.09	&	14:57:28.35	&	0.01	&	6.6	\\
2004.01.11	&	21:31:39.70	&	0.005	&	6.5	&	2006.09.20	&	13:53:39.81	&	1.9	&	11.3	\\
2004.02.03	&	11:07:15.19	&	0.015	&	6.2	&	2006.09.21	&	14:46:12.70	&	0.15	&	5.1	\\
2004.02.14	&	08:06:14.71	&	0.1	&	6.1	&	2006.10.28	&	21:07:51.46	&	0.5	&	5.7	\\
2004.03.17	&	02:13:43.09	&	0.2	&	5.1	&	2007.01.17	&	22:50:42.54	&	15	&	7.1	\\
2004.03.19	&	07:23:22.11	&	0.01	&	6.2	&	2007.01.25	&	14:52:18.42	&	0.015	&	6.4	\\
2004.03.20	&	07:18:39.46	&	0.15	&	5.9	&	2007.03.02	&	01:48:55.43	&	0.01	&	7.2	\\
2004.04.14	&	17:46:41.42	&	0.015	&	6.5	&	2007.03.08	&	19:17:56.92	&	0.15	&	5.9	\\
2004.04.18	&	11:48:08.95	&	0.1	&	5.4	&	2007.03.15	&	22:33:54.91	&	0.2	&	5.3	\\
2004.06.03	&	07:31:40.61	&	0.02	&	6.1	&	2007.06.23	&	09:58:18.97	&	0.015	&	6.3	\\
2004.06.10	&	23:56:16.81	&	0.005	&	6.1	&	2007.07.05	&	21:25:15.50	&	0.01	&	6.2	\\
2004.07.10	&	08:38:34.98	&	0.01	&	7.1	&	2007.08.07	&	04:33:55.41	&	7.5	&	4.7	\\
2004.07.23	&	18:02:03.13	&	0.2	&	5.2	&	2007.09.15	&	12:05:36.20	&	0.02	&	6.9	\\
2004.08.22	&	00:09:03.94	&	0.02	&	6.6	&	2007.09.17	&	08:58:08.22	&	0.01	&	6.9	\\
2004.08.26	&	03:01:44.96	&	0.01	&	6.1	&	2007.09.19	&	21:20:32.48	&	0.02	&	5.3	\\
2004.10.17	&	06:38:49.39	&	1.1	&	5.3	&	2007.09.24	&	01:49:20.51	&	0.005	&	6.2	\\
2004.11.19	&	05:48:28.67	&	0.005	&	6.1	&	2007.10.01	&	01:27:53.33	&	0.1	&	5.8	\\
2004.12.21	&	01:40:04.92	&	0.3	&	5.1	&	2007.10.01	&	03:52:04.94	&	0.005	&	6.3	\\
2004.12.27	&	21:30:26.95	&	0.4	&	8.8	&	2007.10.21	&	05:34:58.21	&	0.01	&	6.6	\\
2005.01.13	&	09:32:22.92	&	0.1	&	5.3	&	2007.11.02	&	03:39:09.38	&	0.01	&	6.6	\\
2005.03.12	&	07:49:36.90	&	0.15	&	5.5	&	2007.11.03	&	01:11:20.07	&	0.01	&	6.6	\\
2005.03.21	&	16:07:18.41	&	0.005	&	6.1	&	2007.12.05	&	12:44:19.70	&	0.005	&	6	\\

\hline

\end{tabular}
\end{table}
\clearpage

\begin{table}[t]

\vspace{6mm} \centering {{\bf Table 7.} Contd.}\label{meansp}

\vspace{5mm}\begin{tabular}{c|c|c|c|c|c|c|c} \hline\hline
Date	&	Time	&	$T_{90}$	      &	Significance	&	Date	&	Time	&	$T_{90}$	      &	Significance\\
        &  UTC      & (s)              &       &       &   UTC  & (s)    &      \\\hline

2007.12.17	&	05:10:35.64	&	0.02	&	6.5	&	2008.10.27	&	22:59:54.53	&	0.05	&	6.5	\\
2007.12.25	&	05:12:22.79	&	0.02	&	6.5	&	2008.11.04	&	03:32:48.82	&	1.9	&	5.1	\\
2007.12.30	&	07:32:05.45	&	0.01	&	6.2	&	2008.11.09	&	00:36:16.55	&	0.005	&	6	\\
2007.12.31	&	18:53:28.50	&	0.005	&	6.1	&	2008.11.16	&	16:54:50.48	&	5	&	4.1	\\
2008.02.23	&	14:05:58.06	&	0.005	&	6.6	&	2008.11.19	&	01:39:21.59	&	0.1	&	5.5	\\
2008.02.24	&	13:06:46.28	&	0.01	&	6.6	&	2008.12.01	&	19:13:22.04	&	0.005	&	6.5	\\
2008.03.13	&	14:54:48.26	&	0.01	&	6.2	&	2008.12.11	&	13:49:31.73	&	0.02	&	7.2	\\
2008.03.27	&	13:29:29.92	&	1.1	&	5.8	&	2008.12.12	&	17:58:51.40	&	0.005	&	6.4	\\
2008.04.08	&	15:34:42.88	&	50	&	44	&	2008.12.12	&	17:59:41.91	&	0.015	&	6.5	\\
2008.04.10	&	14:08:37.39	&	0.05	&	6.3	&	2008.12.14	&	08:55:36.28	&	0.005	&	6.1	\\
2008.04.24	&	00:58:47.19	&	0.005	&	6.7	&	2008.12.15	&	04:56:19.88	&	0.025	&	6.3	\\
2008.04.27	&	21:30:07.39	&	0.02	&	6.3	&	2008.12.19	&	14:56:28.66	&	0.02	&	6.4	\\
2008.04.28	&	07:48:40.79	&	0.01	&	6.3	&	2008.12.23	&	22:00:27.60	&	1.3	&	5.8	\\
2008.04.30	&	22:28:16.60	&	8	&	4.2	&	2008.12.28	&	16:37:26.37	&	0.02	&	7.2	\\
2008.05.15	&	18:51:33.99	&	0.03	&	7.1	&	2009.01.07	&	10:50:03.68	&	0.01	&	7.2	\\
2008.05.23	&	15:33:10.43	&	0.02	&	6.3	&	2009.01.12	&	06:27:43.99	&	0.3	&	5.7	\\
2008.05.27	&	21:27:30.56	&	0.01	&	6.2	&	2009.01.21	&	07:42:19.29	&	0.02	&	6.4	\\
2008.06.05	&	07:27:31.81	&	0.015	&	6.3	&	2009.01.27	&	04:46:14.86	&	0.25	&	5	\\
2008.06.06	&	03:12:30.02	&	0.01	&	6.3	&	2009.03.08	&	10:40:23.59	&	0.02	&	6.5	\\
2008.06.06	&	03:34:29.66	&	0.01	&	7	&	2009.03.13	&	17:10:41.48	&	0.007	&	7.3	\\
2008.06.09	&	02:53:14.45	&	0.02	&	6.3	&	2009.05.24	&	16:07:26.03	&	0.005	&	6.9	\\
2008.06.09	&	10:03:19.56	&	0.2	&	5.6	&	2009.06.09	&	06:26:01.28	&	0.01	&	6.5	\\
2008.06.18	&	09:52:20.43	&	0.015	&	6.3	&	2009.06.14	&	02:15:08.81	&	0.2	&	5.1	\\
2008.06.24	&	15:45:30.81	&	0.15	&	5.1	&	2009.07.03	&	00:52:34.88	&	0.15	&	5.5	\\
2008.07.14	&	03:11:37.75	&	7	&	4.2	&	2009.07.03	&	14:22:27.83	&	0.005	&	7.6	\\
2008.08.03	&	05:25:24.93	&	0.01	&	6.9	&	2009.07.07	&	09:55:01.31	&	0.02	&	6.5	\\
2008.08.03	&	06:23:56.84	&	7	&	4.1	&	2009.07.10	&	19:38:39.98	&	0.7	&	5.4	\\
2008.08.07	&	17:17:25.78	&	0.015	&	6.1	&	2009.07.14	&	08:32:22.24	&	0.015	&	6	\\
2008.08.08	&	03:54:20.00	&	10	&	4.5	&	2009.07.15	&	07:27:04.22	&	1.5	&	5.3	\\
2008.09.16	&	09:26:24.74	&	0.01	&	6.8	&	2009.07.18	&	17:06:03.30	&	0.005	&	6.5	\\
2008.09.23	&	16:48:41.70	&	0.005	&	6.3	&	2009.08.14	&	09:12:40.63	&	0.15	&	5.2	\\
2008.09.28	&	05:20:23.10	&	0.01	&	6.7	&	2009.08.22	&	18:30:26.96	&	0.1	&	5.4	\\
2008.09.30	&	22:39:45.07	&	0.1	&	6.3	&	2009.08.27	&	14:59:49.84	&	0.1	&	5.1	\\
2008.10.03	&	16:23:00.26	&	0.01	&	9.6	&	2009.09.05	&	06:26:26.22	&	0.01	&	6.5	\\
2008.10.03	&	19:23:41.18	&	0.01	&	6.6	&	2009.09.14	&	12:56:56.88	&	1.2	&	5	\\
2008.10.08	&	18:23:33.34	&	0.01	&	6.5	&	2009.09.18	&	01:34:03.62	&	0.01	&	6	\\
2008.10.10	&	07:21:18.65	&	8	&	4.1	&	2009.09.22	&	02:49:17.22	&	0.01	&	6	\\
2008.10.23	&	03:20:11.98	&	0.01	&	6.4	&	2009.10.08	&	01:53:38.54	&	0.1	&	5.4	\\
2008.10.23	&	07:32:53.36	&	0.01	&	6.3	&	2009.11.28	&	06:44:37.75	&	0.25	&	5	\\
2008.10.26	&	00:30:44.97	&	0.01	&	6.8	&	2009.11.29	&	22:17:21.46	&	0.01	&	6.8	\\

\hline

\end{tabular}
\end{table}
\clearpage

\begin{table}[t]

\vspace{6mm} \centering {{\bf Table 8.} Spectral lag of the GRBs detected in the SPI/INTEGRAL experiment determined by the new method}\label{meansp}

\vspace{5mm}\begin{tabular}{c|c|c|c|c|c|c|c} \hline\hline

GRB	&	Telescope	&	N	&	Component	&	Index 1			& Index 2			&	Break	&	$\chi^{2}$ /	dof		\\
	&		&		&		&				&				&	(keV)	&				\\\hline
031203B	&	ISGRI	&	5	&	full$^{g}$	&	0.11	$\pm$	0.60	&	6.15	$\pm$	1.87	&	87.1$\pm$1.2	&	1.66	/	 13	 \\
040223A	&	ISGRI	&	>3$^{f}$	&	full$^{g}$	&	4.66	$\pm$	2.60	&		-		&	-	&	2.82	/	10	\\
	&		&		&	main$^{g}$	&	6.71	$\pm$	2.12	&		-		&	-	&	4.34	/	9	\\
040323A	&	ISGRI	&	>3$^{f}$	&	full$^{g}$	&	3.45	$\pm$	0.47	&		-		&	-	&	1.38	/	13	\\
040422A	&	ISGRI	&	>8$^{f}$	&	full$^{g}$	&	0.32	$\pm$	0.12	&		-		&	-	&	1.17	/	7	\\
040730A	&	ISGRI	&	2	&	full$^{g}$	&	5.11	$\pm$	1.29	&		-		&	-	&	4.43	/	13	\\
040827A	&	ISGRI	&	1	&	full$^{p}$	&	8.51	$\pm$	2.07	&		-		&	-	&	1.34	/	9	\\
041211D	&	SPI	&	1	&	full$^{p}$	&	0.59	$\pm$	0.55	&		-		&	-	&	0.97	/	8	\\
041212A	&	SPI	&	5$^{s}$	&	full$^{g}$	&	5.78	$\pm$	1.97	&	18.41	$\pm$	3.83	&	194.3$\pm$1.2	&	 1.84	/	 10	\\
041218A	&	ISGRI	&	>9$^{v}$	&	full$^{g}$	&	0.21	$\pm$	0.42	&		-		&	-	&	2.61	/	12	\\
	&		&		&	3rd$^{g}$	&	0.09	$\pm$	0.50	&		-		&	-	&	2.30	/	10	\\
050504A	&	ISGRI	&	1	&	full$^{p}$	&	9.83	$\pm$	2.04	&		-		&	-	&	3.34	/	11	\\
050520A	&	ISGRI	&	>7$^{v}$	&	full$^{g}$	&	0.10	$\pm$	0.54	&		-		&	-	&	1.41	/	11	\\
050525A	&	ISGRI	&	3	&	1st$^{g}$*  	&	0.37	$\pm$	0.17	&	-0.36	$\pm$	0.22	&	69.7$\pm$1.2	&	5.85	 /	13	 \\
	&		&		&	2nd$^{p}$* 	&	-0.19	$\pm$	0.09	&		-		&	-	&	1.82	/	13	\\
050525A	&	SPI	&	3	&	1st$^{g}$  	&	0.07	$\pm$	0.13	&		-		&	-	&	0.36	/	7	\\
	&		&		&	2nd$^{p}$  	&	0.52	$\pm$	0.17	&		-		&	-	&	1.06	/	7	\\
	&		&	&	full$^{g}$	&	0.27	$\pm$	0.12	&		-		&	-	&	0.26	/	7	\\
060221C	&	ISGRI	&	>2$^{v}$	&	full$^{g}$	&	0.015	$\pm$	0.023	&		-		&	-	&	0.84	/	10	\\
060428C	&	ISGRI	&	>6$^{v}$	&	full$^{g}$*	&	-0.34	$\pm$	0.13	&		-		&	-	&	1.94	/	10	\\
060428C	&	SPI	&	>6$^{v}$	&	full$^{g}$	&	-1.04	$\pm$	0.97	&	1.71	$\pm$	0.44	&	61.4$\pm$1.2	&	 1.47	/	 14	\\
060901A	&	ISGRI	&	>4$^{v}$	&	full$^{g}$*	&	0.31	$\pm$	0.18	&		-		&	-	&	2.42	/	13	\\
060901A	&	SPI	&	>4$^{v}$	&	full$^{g}$	&	0.11	$\pm$	0.51	&		-		&	-	&	3.29	/	8	\\
061025A	&	ISGRI	&	>2$^{f}$	&	full$^{g}$	&	1.89	$\pm$	0.67	&	-2.77	$\pm$	2.15	&	78.3$\pm$1.3	&	 1.02	 /	9	\\
061122A	&	SPI	&	>3$^{f}$	&	full$^{g}$	&	0.52	$\pm$	0.05	&		-		&	-	&	5.48	/	19	\\
070311A	&	ISGRI	&	1	&	full$^{p}$	&	6.02	$\pm$	1.25	&		-		&	-	&	4.00	/	13	\\
070707B	&	ISGRI	&	3	&	full$^{g}$	&	-0.057	$\pm$	0.032	&		-		&	-	&	1.96	/	10	\\
070912A	&	SPI	&	2	&	full$^{g}$	&	2.40	$\pm$	1.32	&		-		&	-	&	4.53	/	6	\\
070912A	&	SPI+JEM-X	&	2	&	full$^{g}$	&	3.12	$\pm$	1.11	&		-		&	-	&	5.54	/	7	\\
070925C	&	SPI	&	>2$^{v}$	&	full$^{g}$	&	1.63	$\pm$	0.22	&		-		&	-	&	2.35	/	8	\\
071003A	&	SPI	&	>4$^{v}$	&	full$^{g}$	&	-0.20	$\pm$	0.39	&		-		&	-	&	1.65	/	9	\\
080223A	&	ISGRI	&	2	&	full$^{g}$	&	0.60	$\pm$	0.72	&		-		&	-	&	1.76	/	10	\\
080414B	&	ISGRI	&	2	&	1st$^{p}$  	&	0.07	$\pm$	0.18	&		-		&	-	&	1.26	/	12	\\
	&		&		&	full$^{g}$	&	0.14	$\pm$	0.15	&		-		&	-	&	2.71	/	12	\\
	&		&		&	2nd$^{p}$  	&	0.49	$\pm$	0.16	&		-		&	-	&	8.67	/	13	\\
080723B	&	SPI	&	>28$^{f}$	&	1st$^{g}$ 	&	0.44	$\pm$	0.07	&		-		&	-	&	6.58	/	20	\\
	&		&		&	2nd$^{g}$  	&	0.40	$\pm$	0.08	&		-		&	-	&	6.66	/	20	\\
	&		&		&	full$^{g}$	&	0.95	$\pm$	0.24	&	0.11	$\pm$	0.17	&	93.0$\pm$1.3	&	 1.63	/	 16	 \\\hline

\end{tabular}
\end{table}

\begin{table}[t]

\vspace{6mm} \centering {{\bf Table 8.} Contd.}\label{meansp}

\vspace{5mm}\begin{tabular}{c|c|c|c|c|c|c|c} \hline\hline
GRB	&	Telescope	&	N	&	Component	&	Index 1			&	Index 2			&	Break	&	$\chi^{2}$ /	dof		\\
	&		&		&		&				&				&	(keV)	&				\\\hline
081016A	&	SPI	&	5	&	1st$^{g}$  	&	0.32	$\pm$	0.10	&		-		&	-	&	1.67	/	13	\\
	&		&		&	2nd$^{g}$  	&	0.48	$\pm$	0.24	&		-		&	-	&	2.46	/	12	\\
	&		&		&	full$^{g}$	&	0.22	$\pm$	0.18	&		-		&	-	&	2.54	/	13	\\
081226D	&	ISGRI	&	1	&	full$^{p}$	&	0.002	$\pm$	0.002	&		-		&	-	&	3.20	/	15	\\
090625B	&	ISGRI	&	3	&	full$^{g}$	&	1.74	$\pm$	0.37	&		-		&	-	&	2.07	/	11	\\\hline
\multicolumn{8}{l}{* - The study was carried out for part of the event due to the absence of telemetry,}\\
\multicolumn{8}{l}{$^{f}$ - the number of identified and fitted pulses,}\\
\multicolumn{8}{l}{$^{v}$ - the minimum number of pulses estimated visually,}\\
\multicolumn{8}{l}{$^{s}$ - the pulseswere identified using the the SPI-ACS data,}\\
\multicolumn{8}{l}{$^{g}$ - several pulses are analyzed simultaneously,}\\
\multicolumn{8}{l}{$^{p}$ - one pulse is analyzed.}\\

\end{tabular}
\end{table}

\begin{table}[t]

\vspace{6mm} \centering {{\bf Table 9.} Spectral lag of the GRBs detected in the SPI/INTEGRAL experiment between the energy channels 1 - (25, 50), 2 - (50, 100), 3 - (100, 300), and 4 - (300, 1000) keV}\label{meansp}

\vspace{5mm}\begin{tabular}{c|c|c|c|c|c} \hline\hline
GRB	&	Telescope	&	Component	&	lag 2-1	&	lag 3-1	&	lag 4-1	\\
	&		&		&	(s)	&	(s)	&	(s)		\\\hline
031203B	&	IBIS/ISGRI	&	full	&	-0.34$\pm$0.21	&	0.60$\pm$0.26	&	-	\\
040223A	&	IBIS/ISGRI	&	full	&	1.64$\pm$1.43	&	4.31$\pm$1.93	&	-	\\
	&		&	main peak	&	1.77$\pm$1.18	&	3.37$\pm$1.90	&	-	\\
040323A	&	IBIS/ISGRI	&	full	&	1.02$\pm$0.26	&	2.26$\pm$0.32	&	-	\\
040422A	&	IBIS/ISGRI	&	full	&	0.06$\pm$0.04	&	0.04$\pm$0.04	&	-	\\
040730A	&	IBIS/ISGRI	&	full	&	2.31$\pm$0.65	&	2.86$\pm$0.89	&	-	\\
040827A	&	IBIS/ISGRI	&	full	&	3.15$\pm$0.70	&	-	&	-	\\
041218A	&	IBIS/ISGRI	&	3rd peak	&	0.29$\pm$0.22	&	0.79$\pm$0.33	&	-	\\
041218A	&	IBIS/ISGRI	&	full	&	0.19$\pm$0.14	&	0.34$\pm$0.18	&	-	\\
050502A	&	SPI	&	full	&	-0.23$\pm$0.64	&	0.86$\pm$0.47	&	-	\\
050504A	&	IBIS/ISGRI	&	full	&	2.34$\pm$0.77	&	5.64$\pm$1.17	&	-	\\
050520A	&	IBIS/ISGRI	&	full	&	-0.01$\pm$0.30	&	0.67$\pm$0.61	&	-	\\
050525A	&	SPI	&	1st peak	&	0.04$\pm$0.06	&	0.05$\pm$0.07	&	-	\\
050525A	&	IBIS/ISGRI	&	1st peak*	&	0.074$\pm$0.040	&	0.003$\pm$0.053	&	-	\\
050525A	&	SPI	&	2nd peak	&	0.18$\pm$0.12	&	0.25$\pm$0.14	&	-	\\
050525A	&	IBIS/ISGRI	&	2nd peak*	&	0.13$\pm$0.07	&	0.10$\pm$0.07	&	-	\\
050525A	&	SPI	&	full	&	0.10$\pm$0.06	&	0.11$\pm$0.06	&	-	\\
051105A	&	ISGRI	&	full	&	-1.84$\pm$0.79	&	0.32$\pm$0.74	&	-	\\
060221C	&	IBIS/ISGRI	&	full	&	0.008$\pm$0.014	&	0.014$\pm$0.016	&	-	\\
060306C	&	SPI	&	full	&	-0.01$\pm$0.09	&	0.14$\pm$0.43	&	-	\\
060428A	&	SPI	&	full	&	-0.18$\pm$0.19	&	0.22$\pm$0.21	&	-	\\
060428C	&	IBIS/ISGRI	&	full*	&	-0.12$\pm$0.06	&	0.005$\pm$0.077	&	-	\\
060901A	&	SPI	&	full	&	-0.22$\pm$0.30	&	-0.21$\pm$0.31	&	-	\\
060901A	&	IBIS/ISGRI	&	full*	&	-0.09$\pm$0.10	&	0.19$\pm$0.11	&	-	\\
061025A	&	IBIS/ISGRI	&	full	&	0.38$\pm$0.19	&	0.87$\pm$0.28	&	-	\\
061122A	&	SPI	&	full	&	0.11$\pm$0.04	&	0.30$\pm$0.04	&	-	\\
070311A	&	IBIS/ISGRI	&	full	&	1.22$\pm$0.77	&	4.02$\pm$1.18	&	-	\\
070707B	&	IBIS/ISGRI	&	full	&	0.002$\pm$0.018	&	-0.05$\pm$0.02	&	-	\\
070912A	&	SPI	&	full	&	0.84$\pm$0.71	&	1.33$\pm$0.80	&	-	\\
070925C	&	SPI	&	full	&	0.25$\pm$0.35	&	0.78$\pm$0.41	&	-	\\
071003A	&	SPI	&	full	&	-0.24$\pm$0.30	&	-0.33$\pm$0.48	&	-0.55$\pm$0.80	\\
080223A	&	IBIS/ISGRI	&	full	&	0.19$\pm$0.31	&	0.14$\pm$0.60	&	-	\\
080414B	&	IBIS/ISGRI	&	1st peak	&	0.05$\pm$0.10	&	0.25$\pm$0.19	&	-	\\
	&		&	2nd peak	&	0.12$\pm$0.12	&	0.10$\pm$0.24	&	-	\\
	&		&	full	&	0.08$\pm$0.06	&	0.17$\pm$0.15	&	-	\\
080723B	&	SPI	&	1st peak	&	0.21$\pm$0.05	&	0.32$\pm$0.05	&	-	\\
	&		&	2nd peak	&	0.22$\pm$0.07	&	0.27$\pm$0.06	&	0.30$\pm$0.09	\\
	&		&	full	&	0.22$\pm$0.05	&	0.35$\pm$0.04	&	0.33$\pm$0.10	\\
081003C	&	SPI	&	full	&	0.14$\pm$0.60	&	0.25$\pm$0.28	&	-1.77$\pm$1.33	\\\hline
\end{tabular}
\end{table}

\begin{table}[t]

\vspace{6mm} \centering {{\bf Table 9.} Contd.}\label{meansp}

\vspace{5mm}\begin{tabular}{c|c|c|c|c|c} \hline\hline
GRB	&	Telescope	&	Component	&	lag 2-1	&	lag 3-1	&	lag 4-1	\\
	&		&		&	(s)	&	(s)	&	(s)		\\\hline
081016A	&	SPI	&	full	&	0.07$\pm$0.11	&	0.16$\pm$0.2	&		\\
	&		&	1st peak	&	0.06$\pm$0.05	&	0.08$\pm$0.06	&	-	\\
	&		&	2nd peak	&	0.07$\pm$0.13	&	0.36$\pm$0.24	&	-	\\
081226D	&	IBIS/ISGRI	&	full	&	0.006$\pm$0.019	&	0.015$\pm$0.019	&	-	\\
090625B	&	IBIS/ISGRI	&	full	&	0.36$\pm$0.17	&	0.61$\pm$0.26	&	-	\\\hline\hline
GRB	&	Telescope	&	Component	&	lag 3-2	&	lag 4-2	&	lag 4-3	\\
	&		&		&	(s)	&	(s)	&	(s)		\\\hline
041211D	&	SPI	&	full	&	-	&	-	&	0.59$\pm$0.48\\
041212A	&	SPI	&	full	&	2.59$\pm$0.61	&	7.54$\pm$0.84	&	-\\
051220A	&	SPI	&	full	&	0.23$\pm$0.49	&	0.36$\pm$0.39	&	-\\
081110C	&	SPI	&	full	&	-	&	-	&	-0.05$\pm$0.06\\\hline
\multicolumn{6}{l}{}\\ [-3mm]
\multicolumn{6}{l}{* - There are no data for part of the event due to telemetry failure.}\\

\end{tabular}
\end{table}

\clearpage

\clearpage

\begin{figure}[t]
\epsfxsize=13cm \hspace{0.5cm} \vspace{0cm} \epsffile{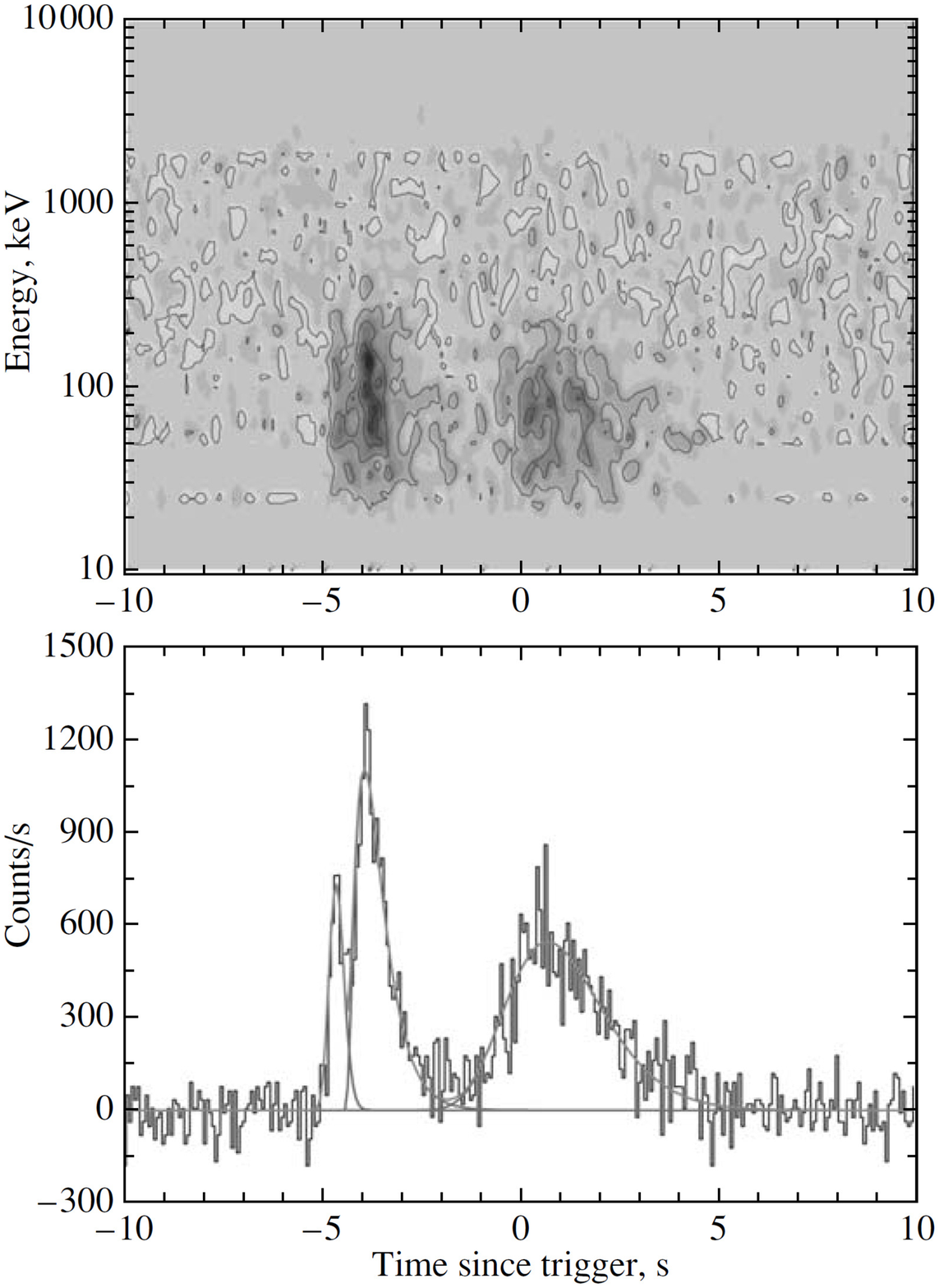}\\

\textbf{Fig.1.} GRB 050525A. The upper panel shows the energy–time diagram constructed from the SPI data. The time since the trigger in seconds is along the horizontal axis. The photon energy in keV is along the vertical axis. The color brightness is inversely proportional to the number of recorded counts in the corresponding region: the darker the region, the larger the number of counts in it. The lower panel shows the light curve in the energy range (20, 2000) keV constructed from the SPI data. The time since the trigger in seconds is along the horizontal axis. The number of counts per second in a light-curve bin is along the vertical axis. The smooth curve indicates the fit to the light curve by the sum of exponential pulses (Eq. (5)).
\end{figure}

\begin{figure}[h]
\epsfxsize=13cm \hspace{0.5cm}\epsffile{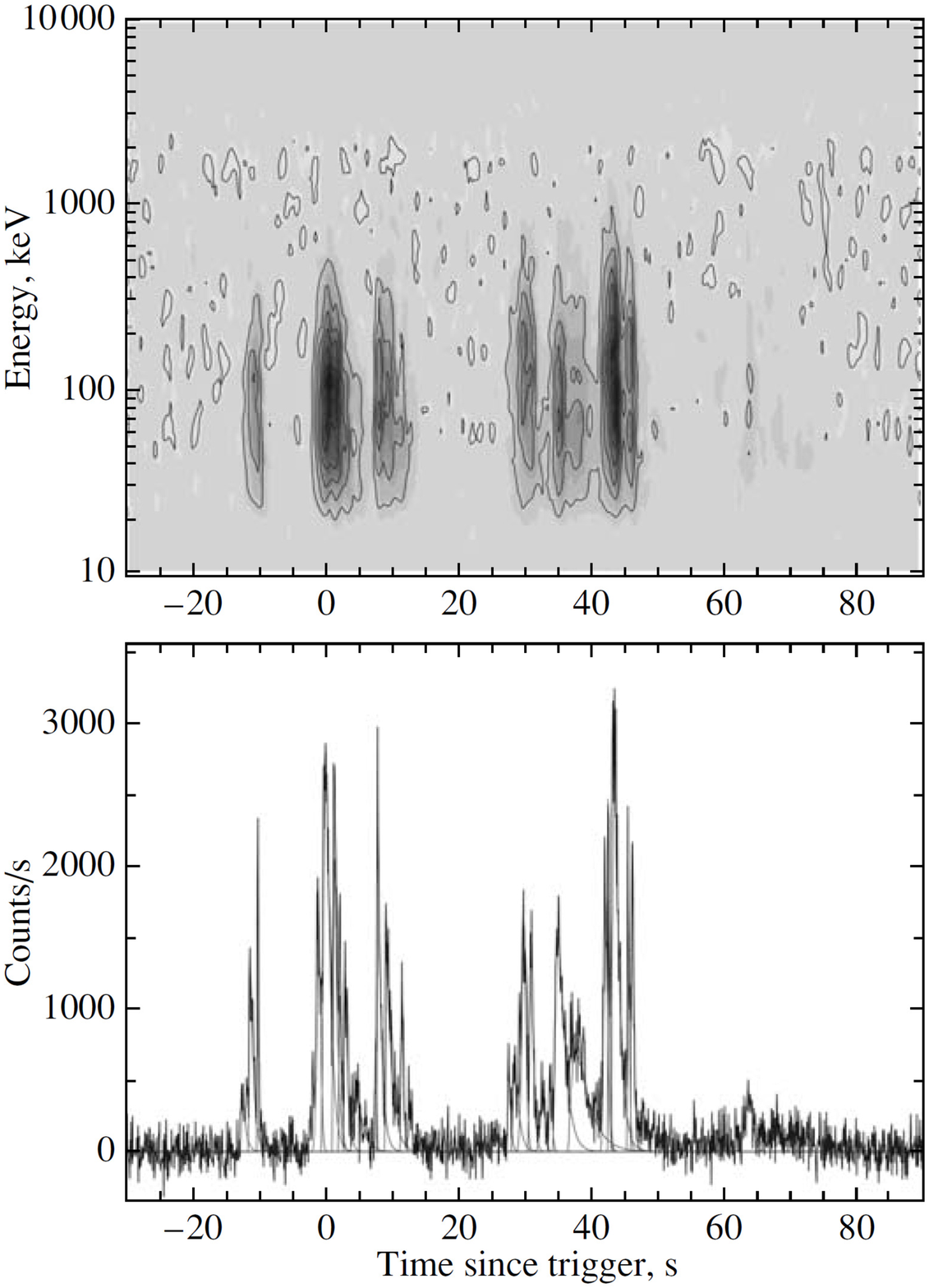}\\

\textbf{Fig.2.} GRB 080723B. The same as Fig. 1.

%\caption{\rm }
\end{figure}

\begin{figure}[h]
\epsfxsize=14cm \hspace{0cm} \epsffile{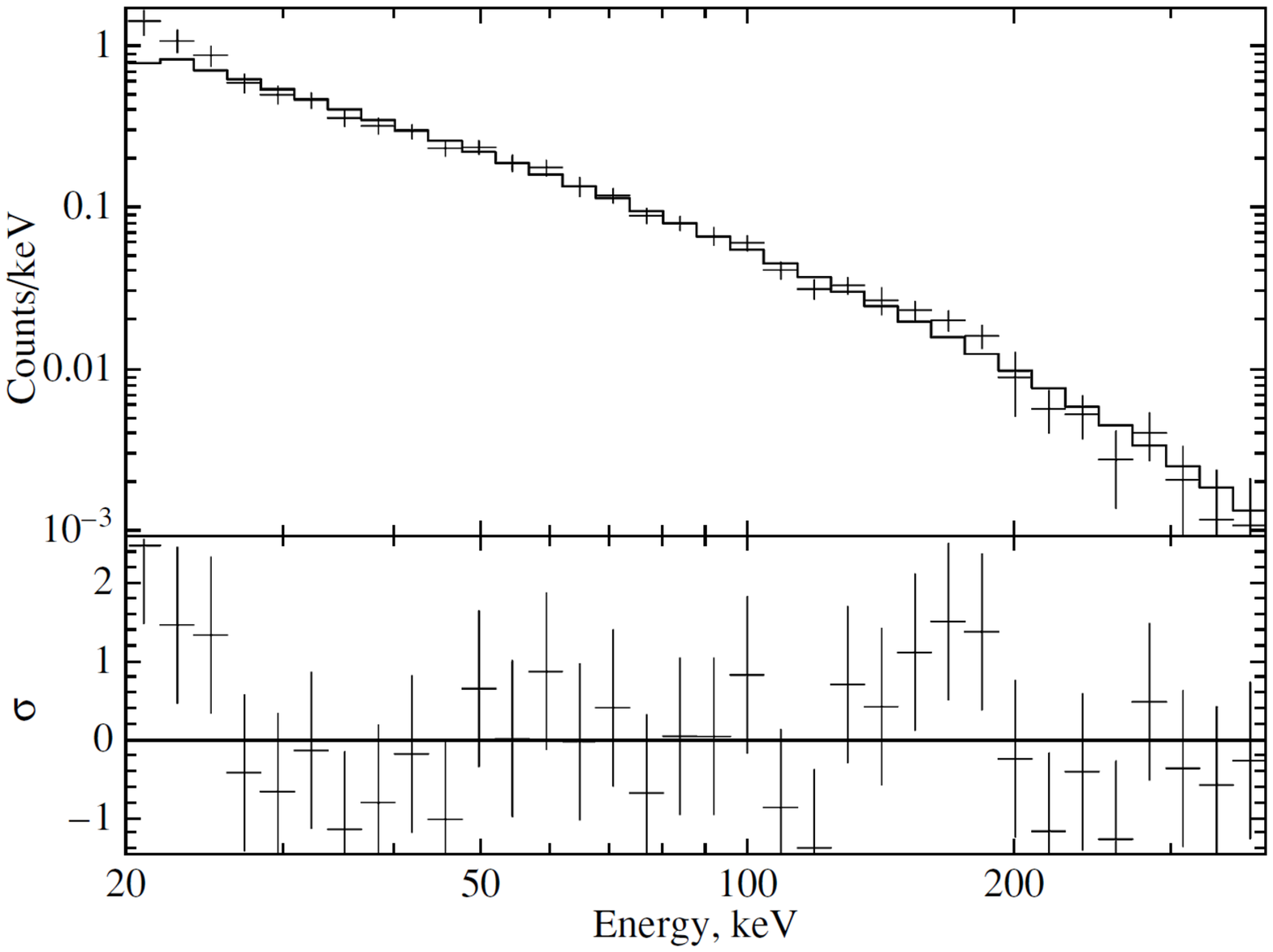}\\

\textbf{Fig.3.} Energy spectrum of GRB 050525A. The photon energy in keV is along the horizontal axis. The number of counts per second per keV is along the vertical axis. The solid line indicates the power-law spectral model with an exponential cutoff (Eq. (3)). In the lower panel, the deviation of the spectral model from the observational data is plotted against the photon energy. The model deviation from the observational data in standard deviations is along the vertical axis.

%\caption{\rm }
\end{figure}

\begin{figure}[h]
\epsfxsize=14cm \hspace{0cm} \epsffile{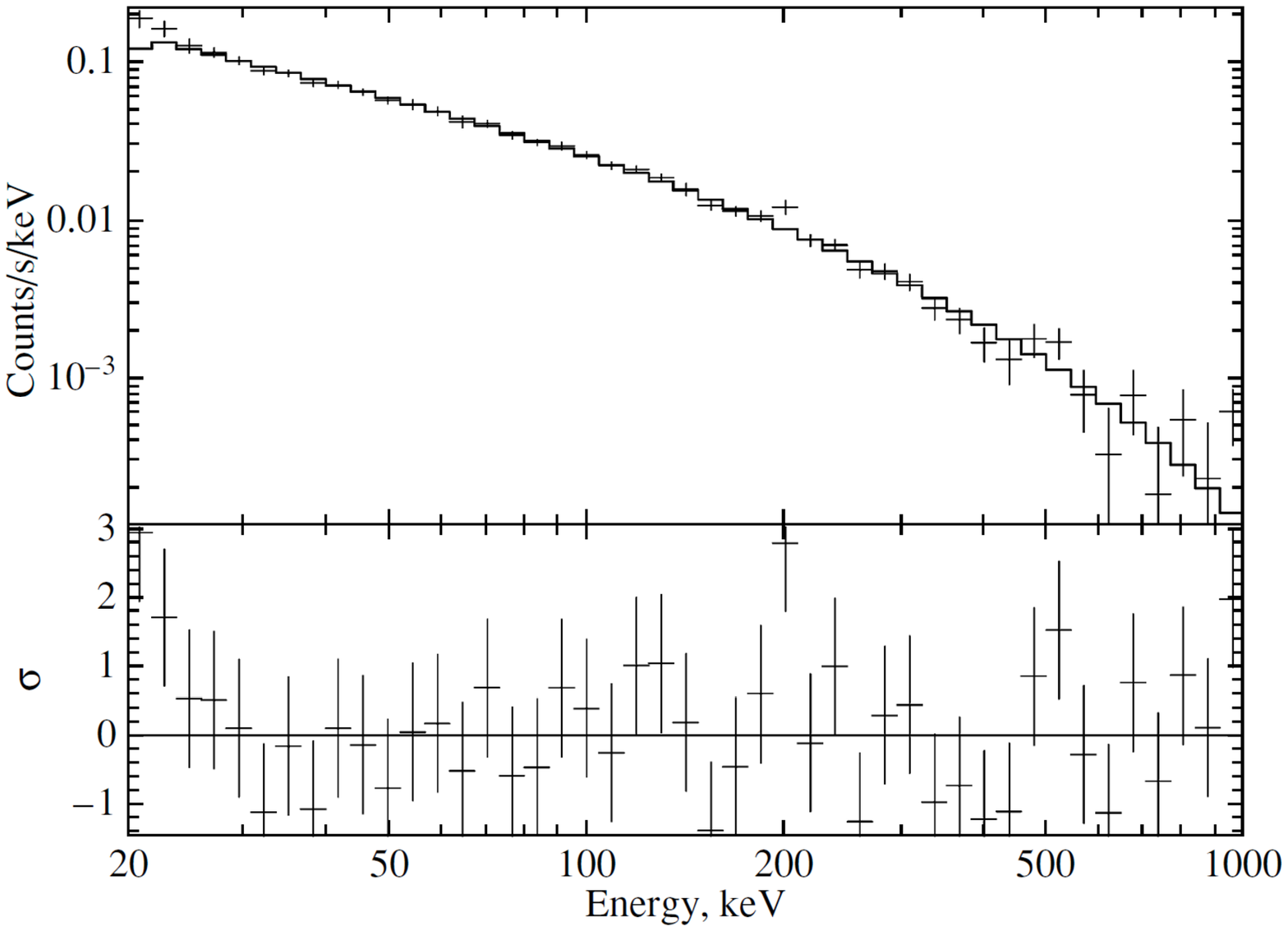}\\

\textbf{Fig.4.} Energy spectrum of GRB 080723B. The same as Fig. 3.
%\caption{\rm }
\end{figure}

\begin{figure}[h]
\epsfxsize=14cm \hspace{0cm} \epsffile{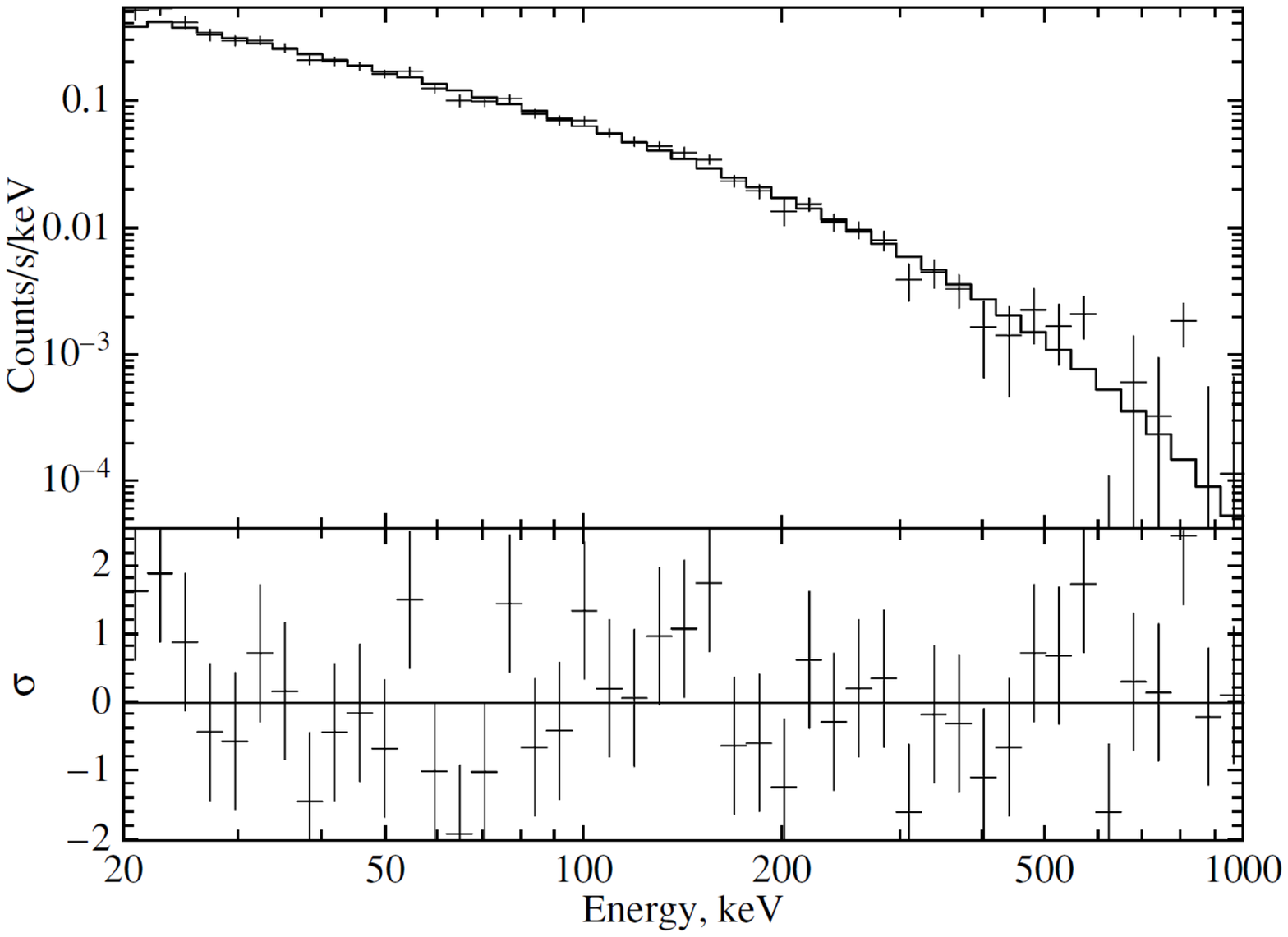}\\

\textbf{Fig.5.} Energy spectrum of GRB 061122A. The same as Fig. 3.
%\caption{\rm }
\end{figure}

\begin{figure}[h]
\epsfxsize=14cm \hspace{0cm} \epsffile{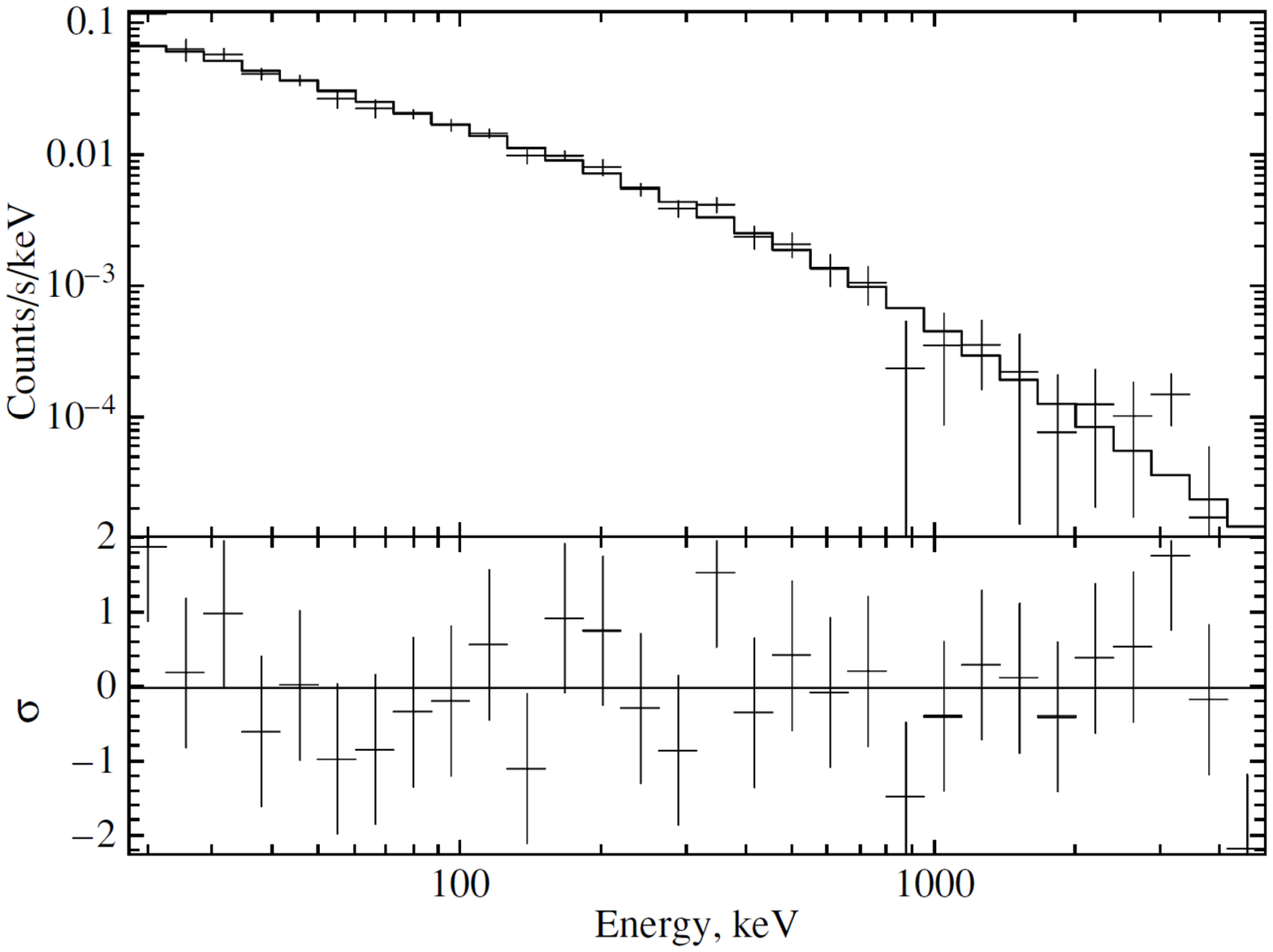}\\

\textbf{Fig.6.} Energy spectrum of GRB 071003A. The photon energy in keV is along the horizontal axis. The number of counts per second per keV is along the vertical axis. The solid line indicates Band’s spectral model (Eq. (4)). In the lower panel, the deviation of the spectral model from the observational data is plotted against the photon energy. The model deviation from the observational data in standard deviations is along the vertical axis.
%\caption{\rm }
\end{figure}

\begin{figure}[h]
\epsfxsize=15cm \hspace{0cm}\epsffile{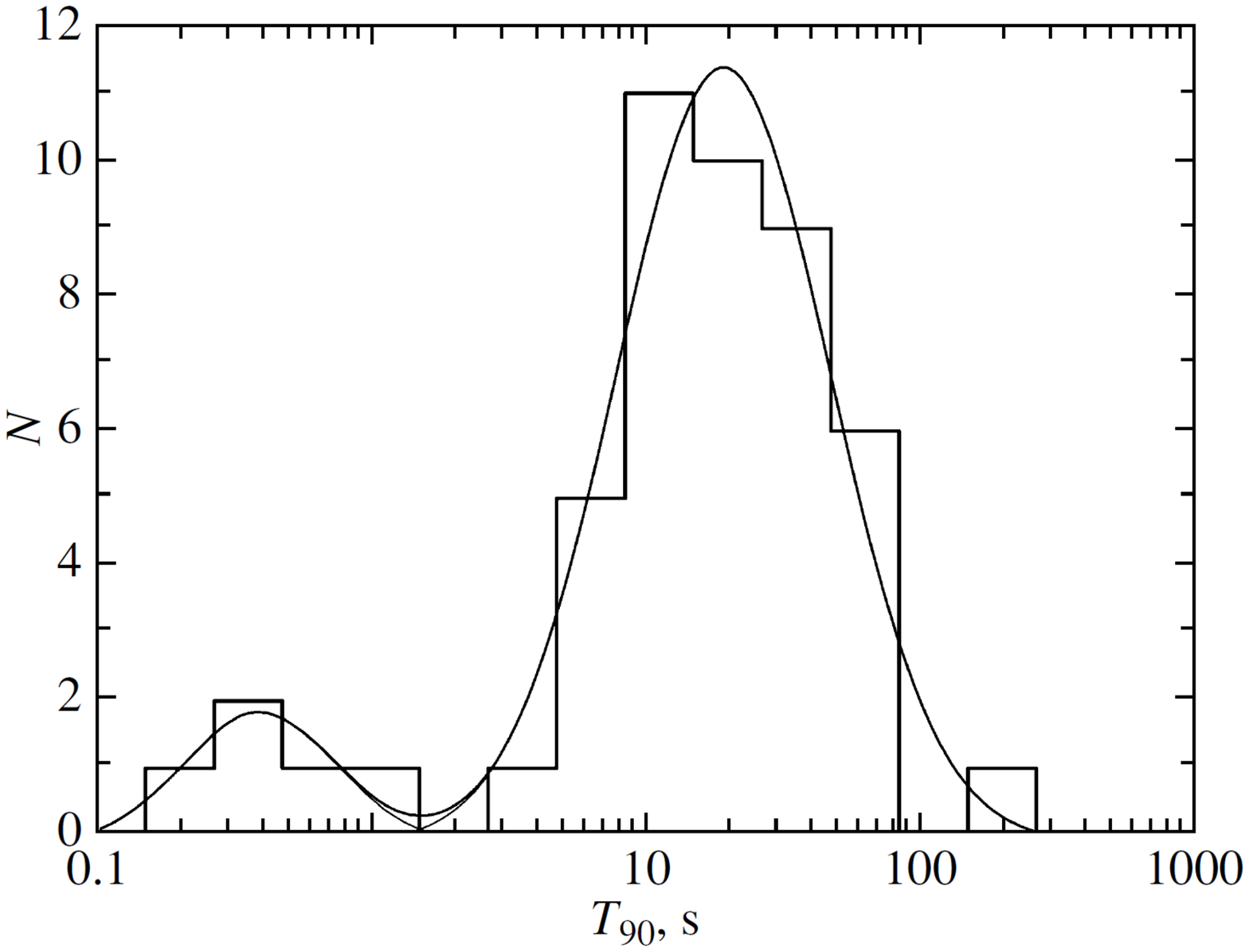}\\

\textbf{Fig.7.} Duration distribution of the GRBs confirmed by other space telescopes in duration \t90. The GRB duration in seconds is along the horizontal axis; the number of bursts is along the vertical axis. The smooth curve indicates the fit to the observed dependence by two log-normal distributions. The centers of the log-normal distributions are located at T$_{90}^{short}$ = 0.39$_{-0.09}^{+0.12}$ s, T$_{90}^{long}$ = 19.23$_{-1.04}^{+1.09}$ s.
%\caption{\rm }
\end{figure}

\begin{figure}[h]
\epsfxsize=13cm \hspace{0cm}\epsffile{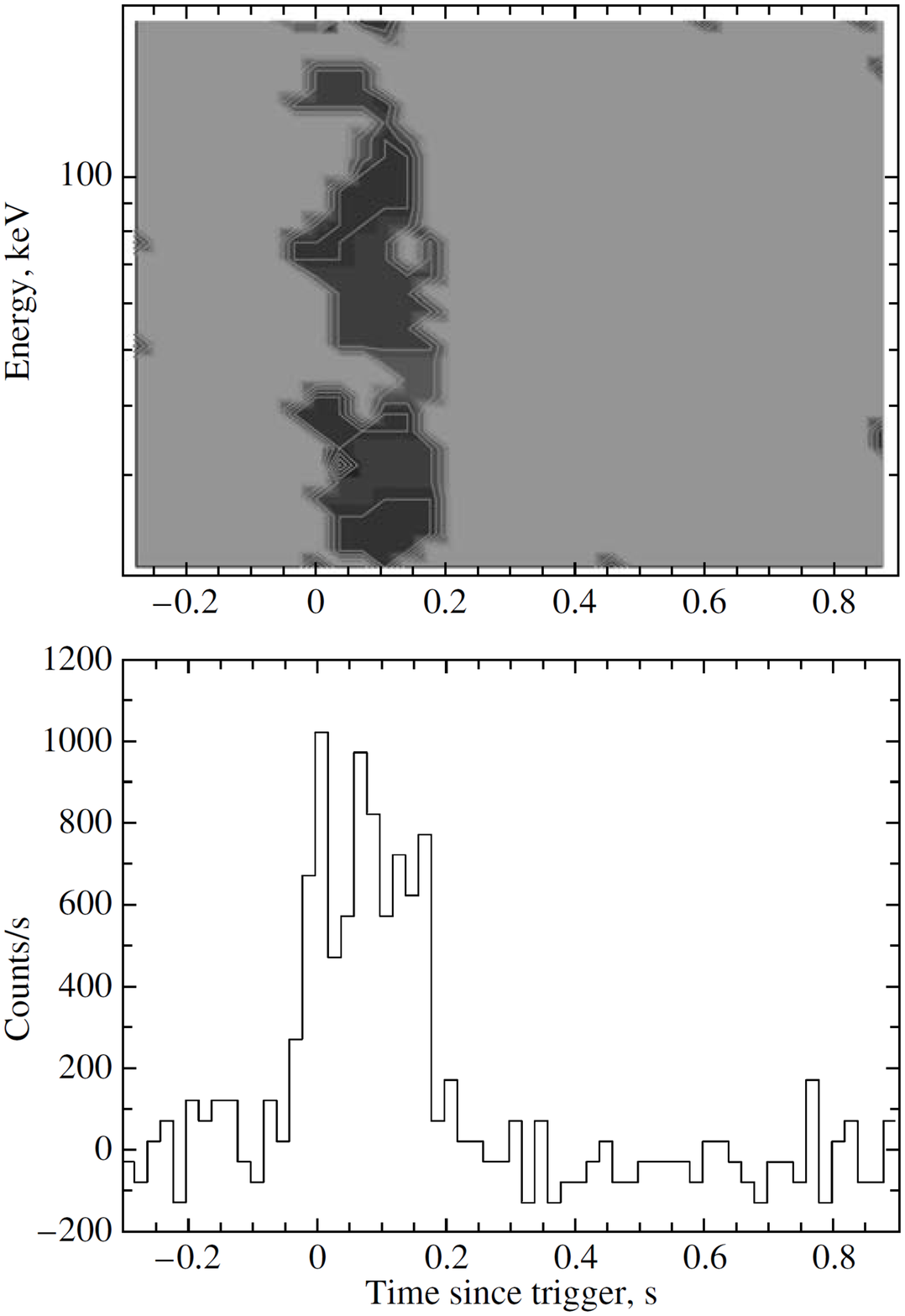}\\

\textbf{Fig.8.} GRB 060221C. The upper panel shows the energy–time diagram constructed from the IBIS/ISGRI data. The time since the trigger in seconds is along the horizontal axis. The photon energy in keV is along the vertical axis. The color brightness is inversely proportional to the number of recorded counts in the corresponding region: the darker the region, the larger the number of counts in it. The lower panel shows the light curve in the energy range (20, 2000) keV constructed from the IBIS/ISGRI data. The time since the trigger in seconds is along the horizontal axis.
%\caption{\rm }
\end{figure}

\begin{figure}[h]
\epsfxsize=15cm \hspace{0cm}\epsffile{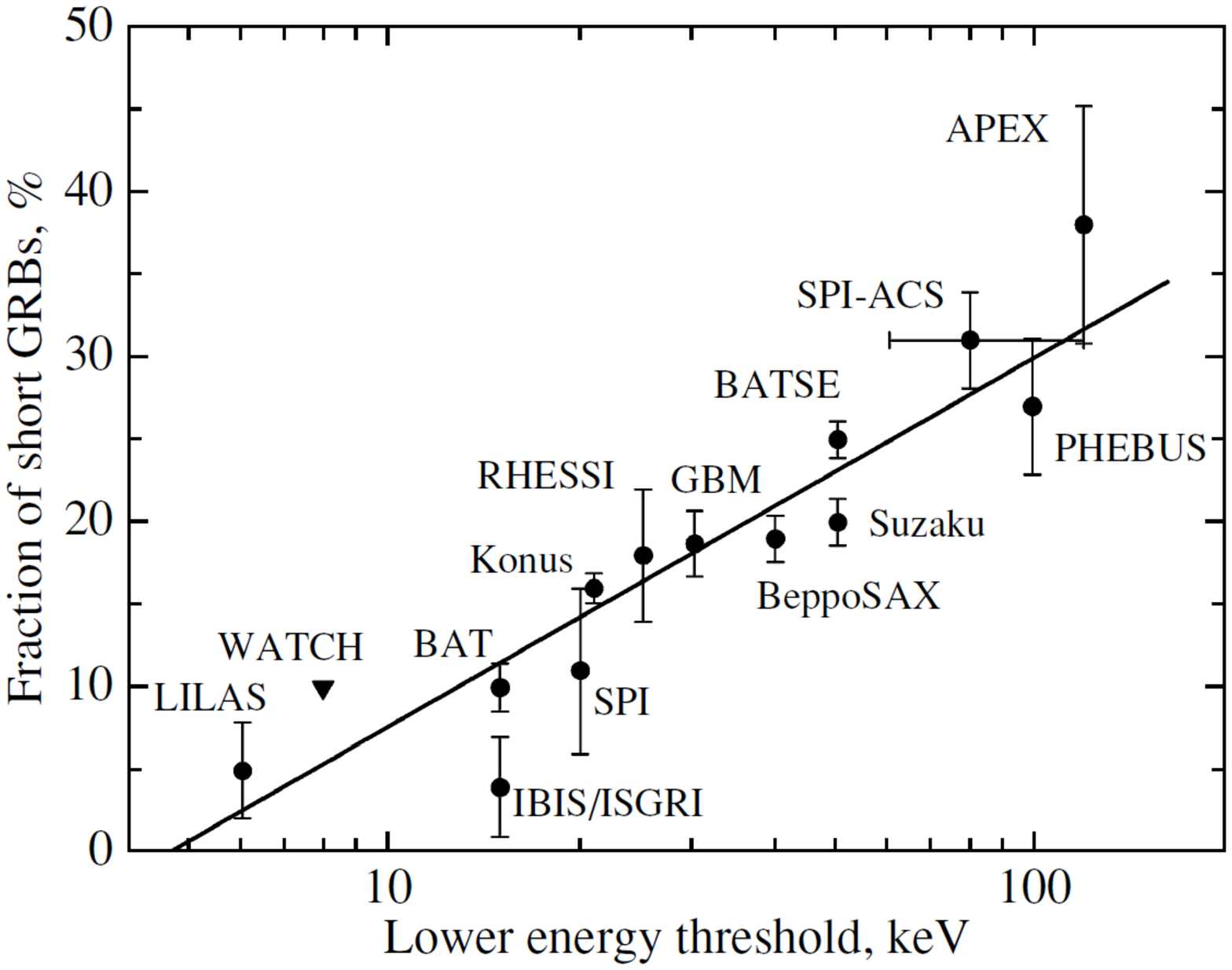}\\

\textbf{Fig.9.} Fraction of short GRBs versus lower energy threshold of the detector trigger algorithm in various experiments. The lower energy threshold of the detector triggers in keV is along the horizontal axis; the fraction of short GRBs in the experiment in percent of the total number of bursts is along the vertical axis. An upper limit for the fraction of short GRBs is indicated for the GRANAT/WATCH experiment (Sazonov et al. 1998).
%\caption{\rm }
\end{figure}

\begin{figure}[h]
\epsfxsize=14cm \hspace{0cm}\epsffile{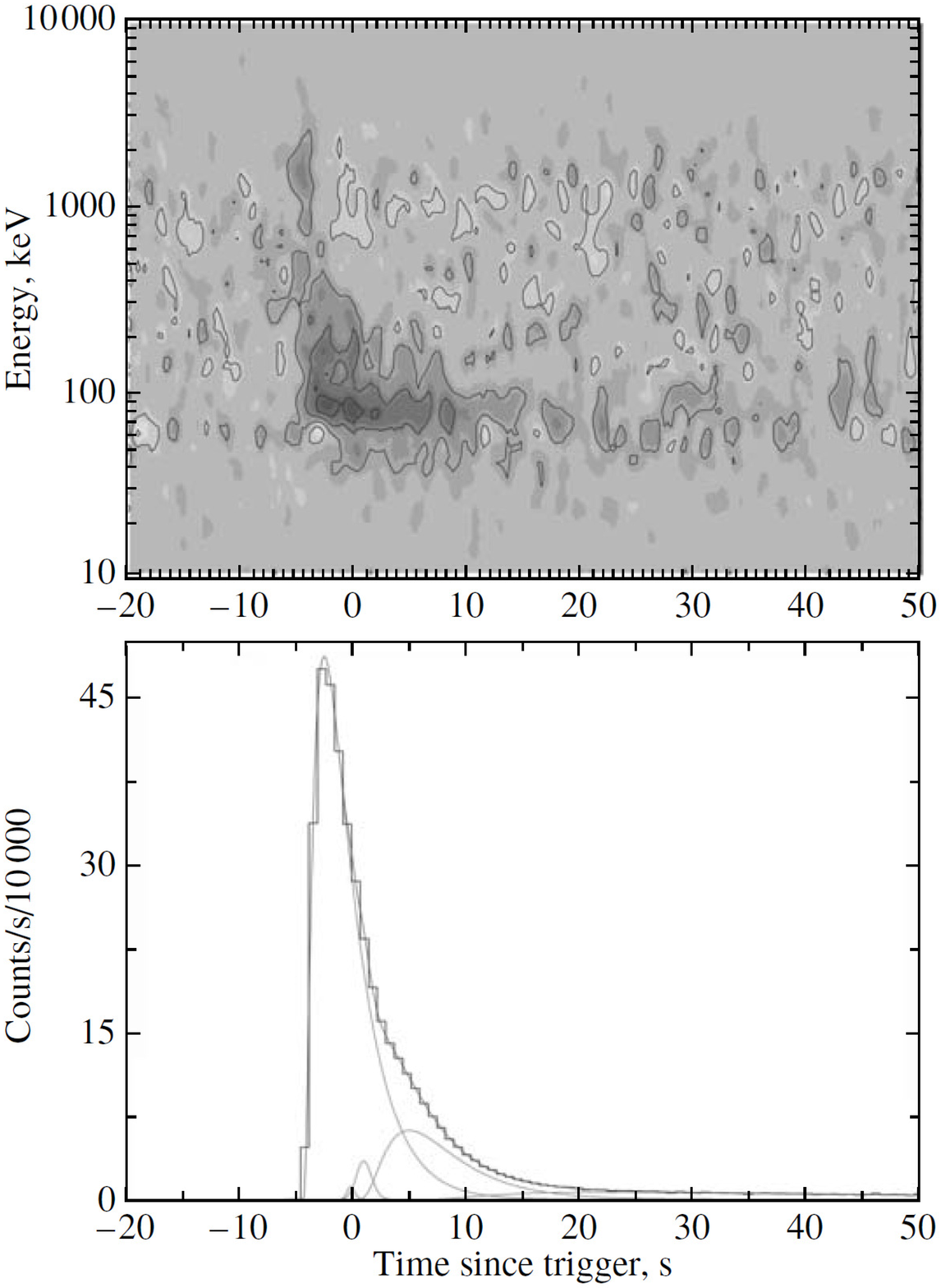}\\

\textbf{Fig.10.} GRB 041212A. The upper panel shows the energy–time diagram constructed from the SPI data. The time since the trigger in seconds is along the horizontal axis. The photon energy in keV is along the vertical axis. The color brightness is inversely proportional to the number of recorded counts in the corresponding region: the darker the region, the larger the number of counts in it. The lower panel shows the light curve in the energy range (20, 2000) keV constructed from the SPI–ACS data. The time since the trigger in seconds is along the horizontal axis. The number of counts per second in a light-curve bin is along the vertical axis. The smooth curve indicates the fit to the light curve by the sum of exponential pulses (Eq. (5)).
%\caption{\rm }
\end{figure}

\begin{figure}[h]
\epsfxsize=15cm \hspace{0cm}\epsffile{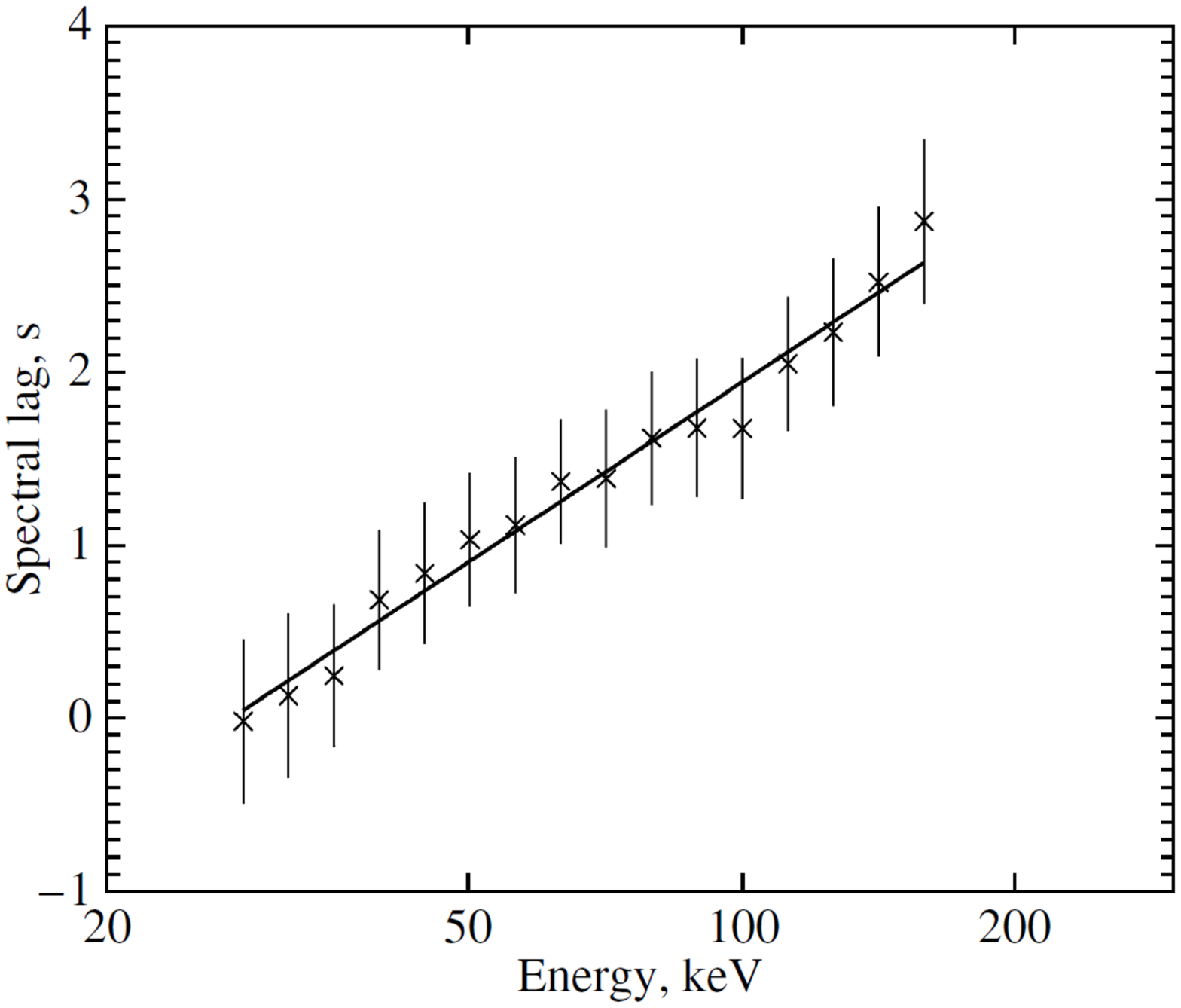}\\

\textbf{Fig.11.} Energy dependence of the spectral lag for GRB 040323A constructed from the IBIS/ISGRI data. The fit to the dependence by a logarithmic function (Eq. (6)) is shown. The energy in keV is along the horizontal axis; the spectral lag in seconds is along the vertical axis.
%\caption{\rm }
\end{figure}

\begin{figure}[h]
\epsfxsize=15cm \hspace{0cm}\epsffile{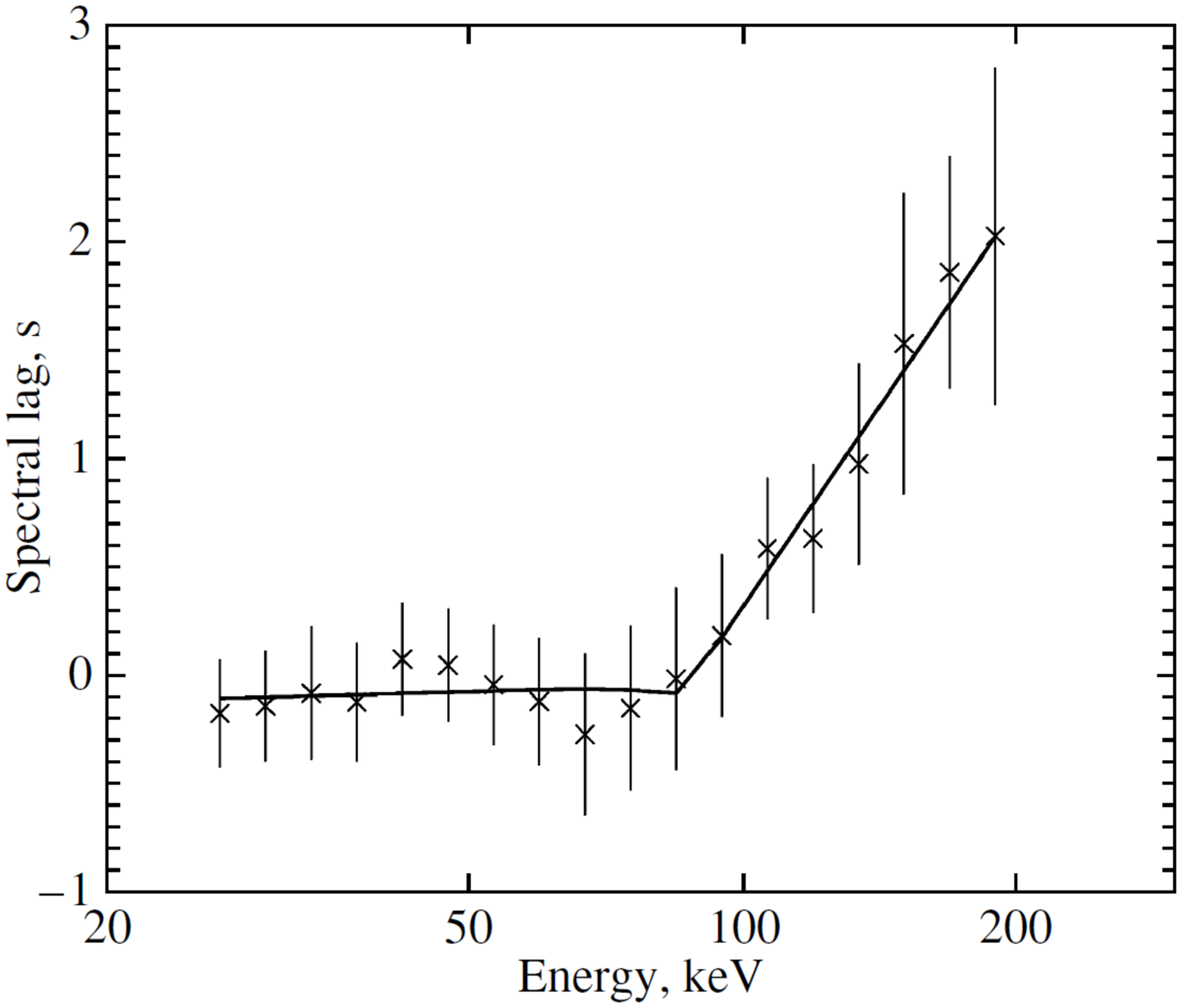}\\

\textbf{Fig.12.} Energy dependence of the spectral lag for GRB 031203B constructed from the IBIS/ISGRI data. The fit to the dependence by a logarithmic function with a break (Eq. (7)) is shown. The energy in keV is along the horizontal axis; the spectral lag in seconds is along the vertical axis.
%\caption{\rm }
\end{figure}

\begin{figure}[h]
\epsfxsize=15cm \hspace{0cm}\epsffile{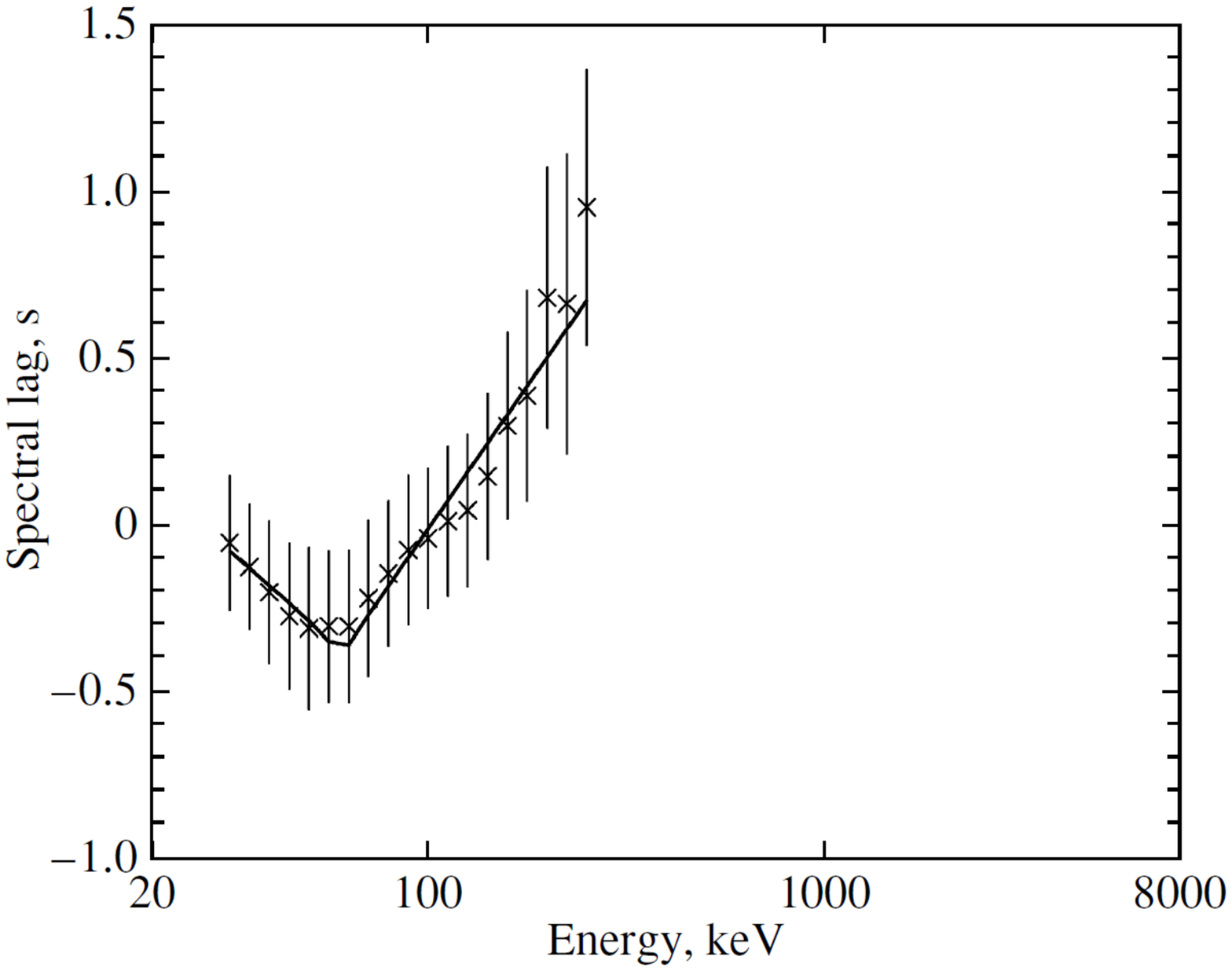}\\

\textbf{Fig.13.} Energy dependence of the spectral lag for GRB 060428C constructed from the SPI data. The same as Fig. 12.
%\caption{\rm }
\end{figure}

\begin{figure}[h]
\epsfxsize=15cm \hspace{0cm}\epsffile{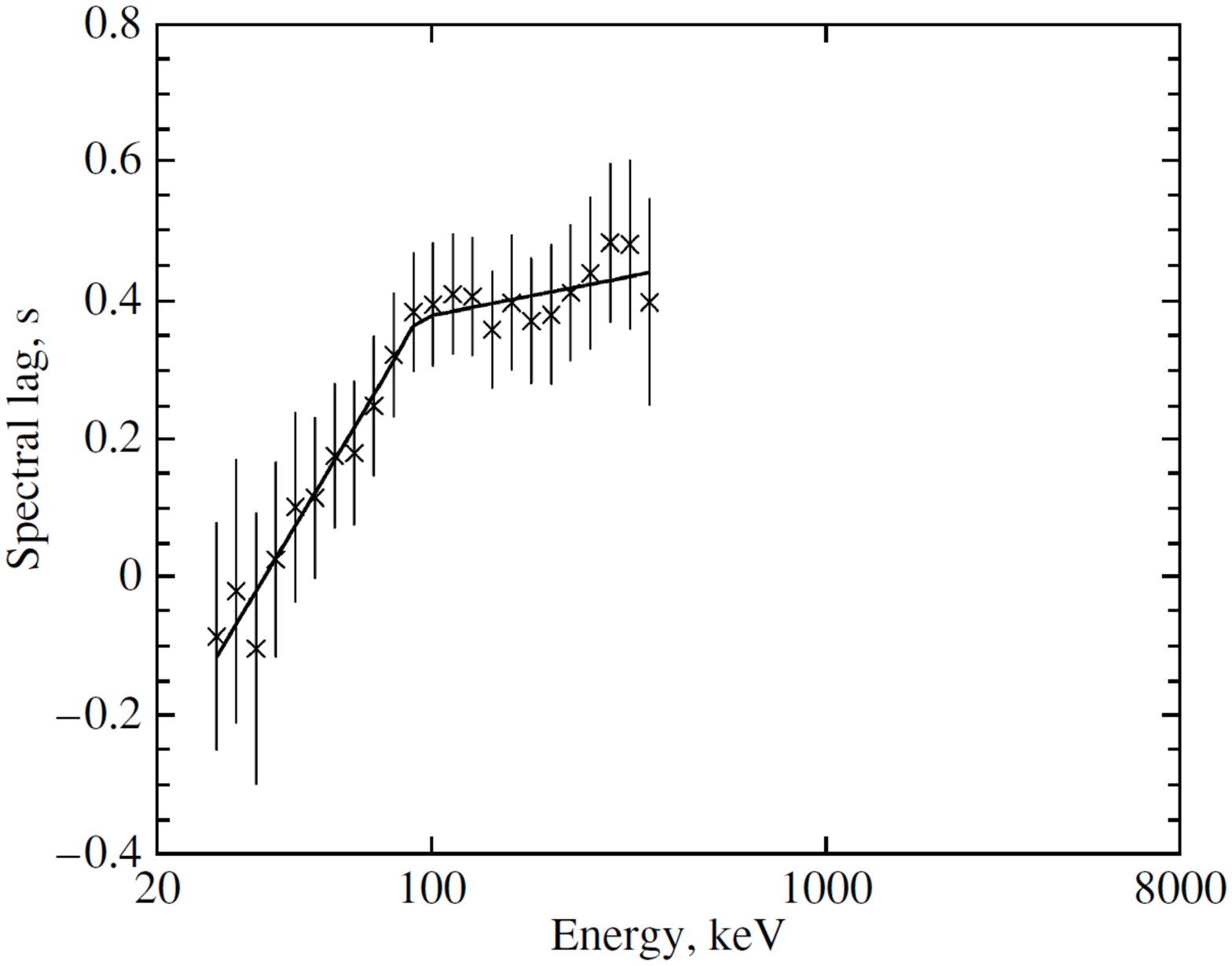}\\

\textbf{Fig.14.} Energy dependence of the spectral lag for GRB080723A constructed from the SPI data. The same as Fig. 12.
%\caption{\rm }
\end{figure}

\begin{figure}[h]
\epsfxsize=15cm \hspace{0cm}\epsffile{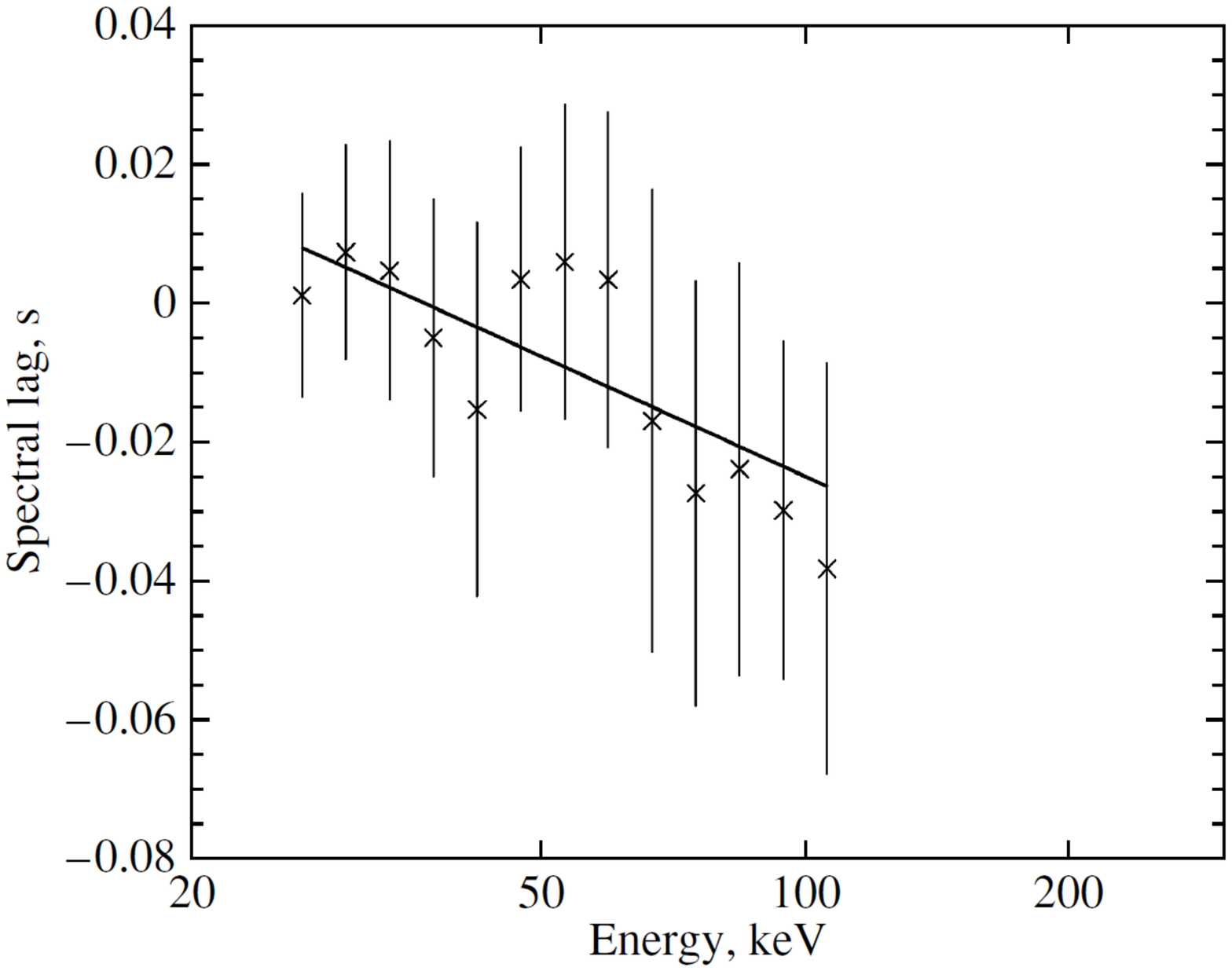}\\

\textbf{Fig.15.} Energy dependence of the spectral lag for GRB 070707B constructed from the IBIS/ISGRI data. The same as in Fig. 11.
%\caption{\rm }
\end{figure}

\begin{figure}[h]
\epsfxsize=15cm \hspace{0cm}\epsffile{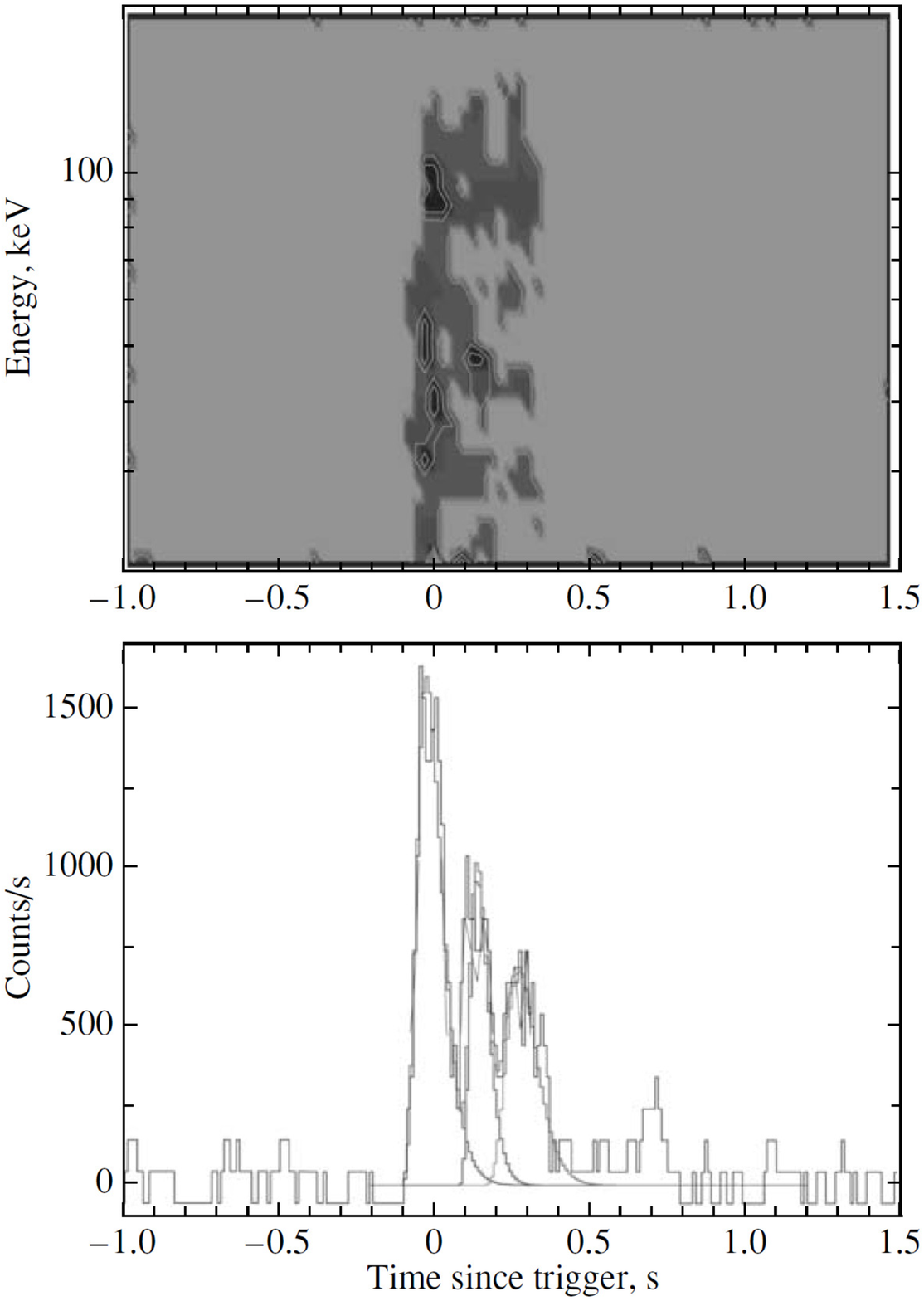}\\

\textbf{Fig.16.} GRB 070707B. The same as in Fig. 8.
%\caption{\rm }
\end{figure}

\begin{figure}[h]
\epsfxsize=14cm \hspace{0cm}\epsffile{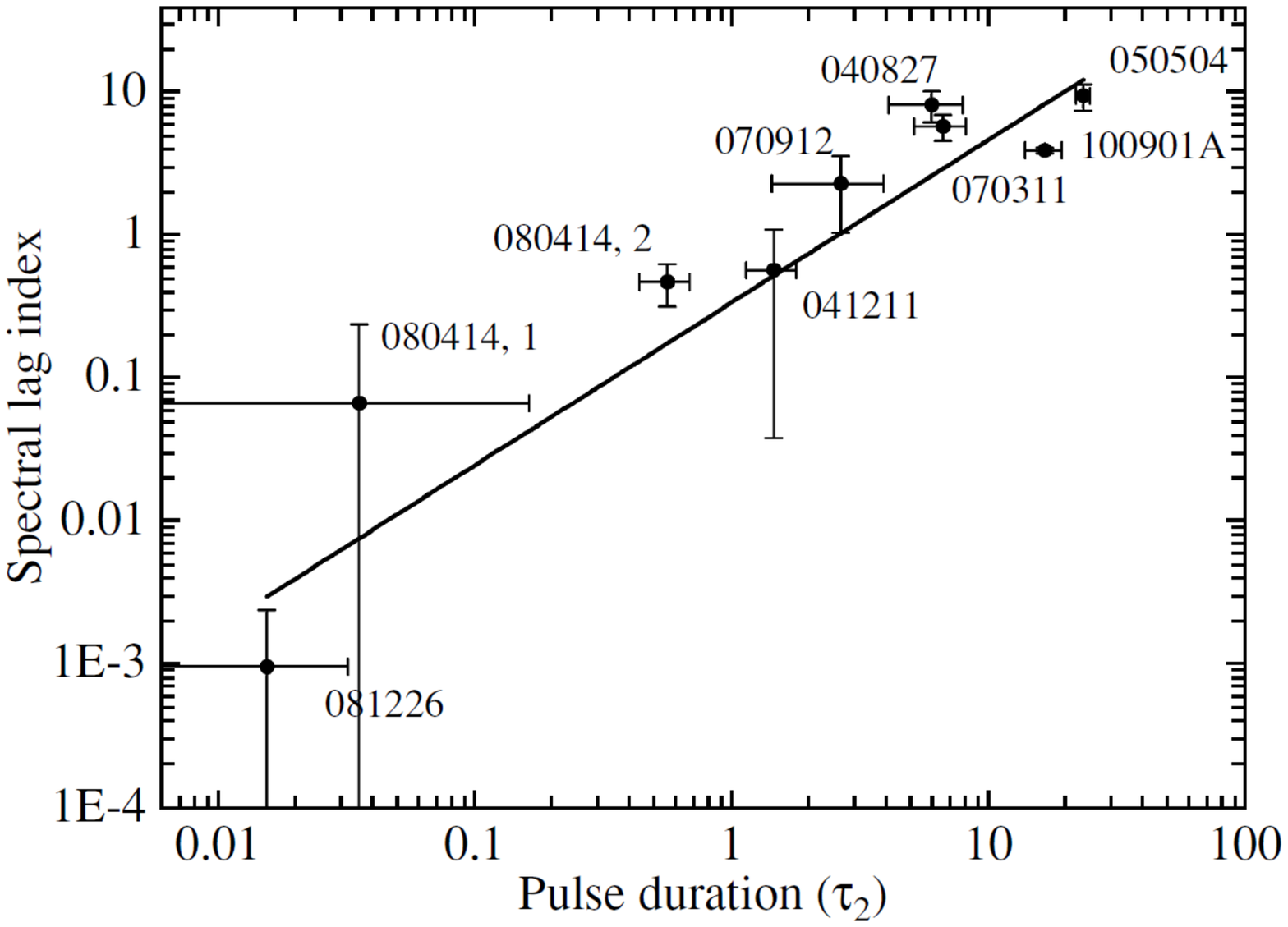}\\

\textbf{Fig.17.} Spectral lag index versus pulse duration. The pulse duration in units of the parameter $\tau_{2}$ is along the horizontal axis. The spectral lag index is along the vertical axis. For each point, we specify what GRB it corresponds to. The straight line indicates the fit to this dependence by a power law with an exponent 1.07 $\pm$ 0.10.
%\caption{\rm }
\end{figure}

\end{document}